\documentclass{aastex63}

\newcommand{\cawav}{\ion{Ca}{2}~8542\,\AA}
\newcommand{\ca}{\ion{Ca}{2}}
\newcommand{\ha}{H$\alpha$}
\newcommand{\hb}{H$\beta$}
\newcommand{\heir}{\ion{He}{1}~10830\,\AA}
\newcommand{\he}{\ion{He}{1}~5876\,\AA\ (D3)}
\newcommand{\cmcub}{cm$^{-3}$}
\newcommand{\kms}{km\,s$^{-1}$}

\usepackage{soul}

\received{MM DD, 2020}
\revised{MM DD, 2020}
\accepted{MM DD, 2020}

\shorttitle{Diagnostics of spicule}
\shortauthors{Kuridze et al.}

\begin{document}

\title{Semi-empirical models of spicule from inversion of \cawav\ line}

\correspondingauthor{D. Kuridze}
\email{dak21@aber.ac.uk}

\author[0000-0003-2760-2311]{David Kuridze} 
\affiliation{Department of Physics, Aberystwyth University, Ceredigion, SY23 3BZ, UK}
\affiliation{Abastumani Astrophysical Observatory, Mount Kanobili, 0301, Abastumani, Georgia}

\author[0000-0001-9896-4622]{Hector Socas-Navarro}
\affiliation{Instituto de Astrof\'{\i}sica de Canarias, 38205, C/Via L\'actea s/n, La Laguna, Tenerife, Spain}
\affiliation{Departamento de Astrof\'{\i}sica, Universidad de La Laguna, La Laguna, E-38205, Tenerife, Spain}

\author[0000-0002-7444-7046]{J\'{u}lius Koza} 
\affil{Astronomical Institute, Slovak Academy of Sciences, 059 60 Tatransk\'{a} Lomnica, Slovakia}

\author[0000-0003-4162-7240]{Ramon Oliver}
\affiliation{Departament de F\'{\i}sica, Universitat de les Illes Balears, E-07122 Palma de Mallorca, Spain}
\affiliation{Institute of Applied Computing \& Community Code (IAC3), UIB, Spain}

\begin{abstract}
We study a solar spicule observed off-limb using high-resolution
imaging spectroscopy in the \cawav\ line obtained with the CRisp Imaging SpectroPolarimeter (CRISP) 
on the Swedish 1-m Solar Telescope. Using a
new version of the non-LTE code NICOLE specifically developed for this
problem we invert the spicule single- and double-peak line
profiles. This new version considers off-limb geometry and
computes atomic populations by solving the 1D radiative transfer
assuming a vertical stratification. The inversion proceeds by
fitting the observed spectral profiles at 14 different heights with
synthetic profiles computed in the model by solving the radiative transfer problem along its length.
Motivated by the appearance of double-peak
\cawav\ spicule profiles, which exhibit two distinct emission features
well separated in wavelength, we adopt a double-component
scenario. We start from the ansatz 
that the spicule parameters are
practically constant along the spicule axis for each component, except
for a density drop. Our results support this ansatz by attaining very
good fits to the entire set of 14\,$\times$\,4 profiles (14 heights and 4 times). 
We show
that the double-component model 
with uniform temperature of 9\,560\,K, exponential decrease of
density with a height scale of $1\,000-2\,000$\,km, and 
the counter-oriented line-of-sight velocities of components
reproduce the double-peak line profiles at all spicule
segments well. Analyses of the numerical response function reveals the necessity of the 
inversions of spectra at multiple height
positions to obtain height-dependent, degeneracy-free reliable model with a limited number of free parameters.

\end{abstract}


\section{Introduction}

Spicules are small-scale, jet-like plasma features observed
ubiquitously at the solar limb and have been 
reviewed by
\citet{Beckers1968,Beckers1972}, \citet{Sterling2000}, and
\citet{Tsiropoula2012}. Originating in the active regions and network
boundaries, they protrude into the corona and can act as conduits
channeling energy and mass from the solar photosphere into the upper
layers. Despite tremendous observational and theoretical efforts since
their discovery in the 19th century \citep{Secchi1975}, many aspects
of spicule physics, such as their formation mechanism, magnetism, and
thermodynamic properties remain the central subject of ongoing solar
research.

\begin{figure}
\includegraphics[width=\textwidth]{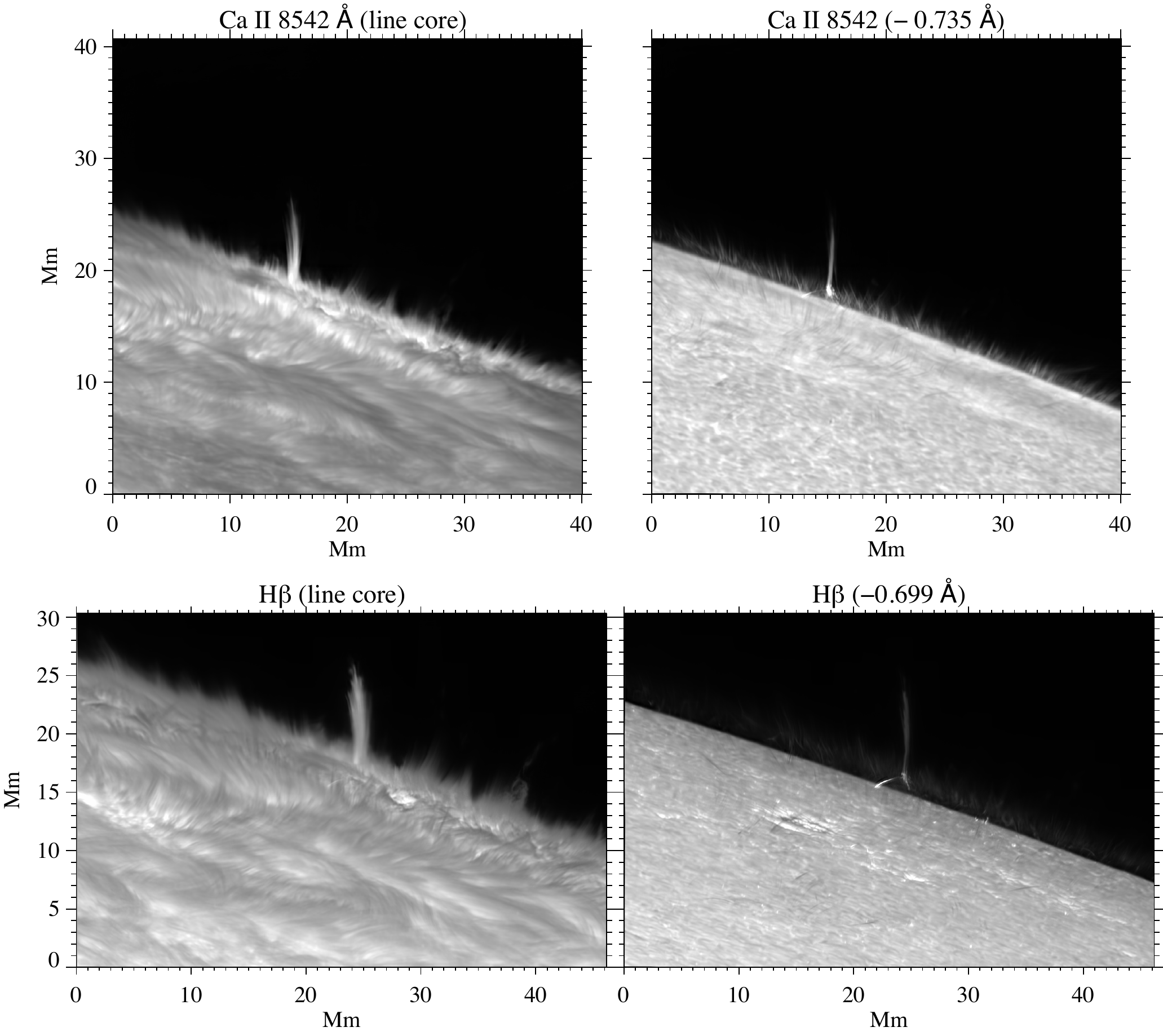}
\caption{Overview of SST observations of the spicule on 2018 June 22
  at 08:34:27\,UT at the western limb near AR NOAA 12714. Each image
  is byte-scaled independently. Top: CRISP \cawav\ line core image
  (left) and blue wing image at $\Delta\lambda =
  -0.735$\,\AA\ (right). Bottom: CHROMIS \hb\ line core image (left)
  and blue wing image at $\Delta\lambda = -0.699$\,\AA\ (right).}
\label{fig1}
\end{figure}

\begin{figure}
\includegraphics[width=0.625\textwidth]{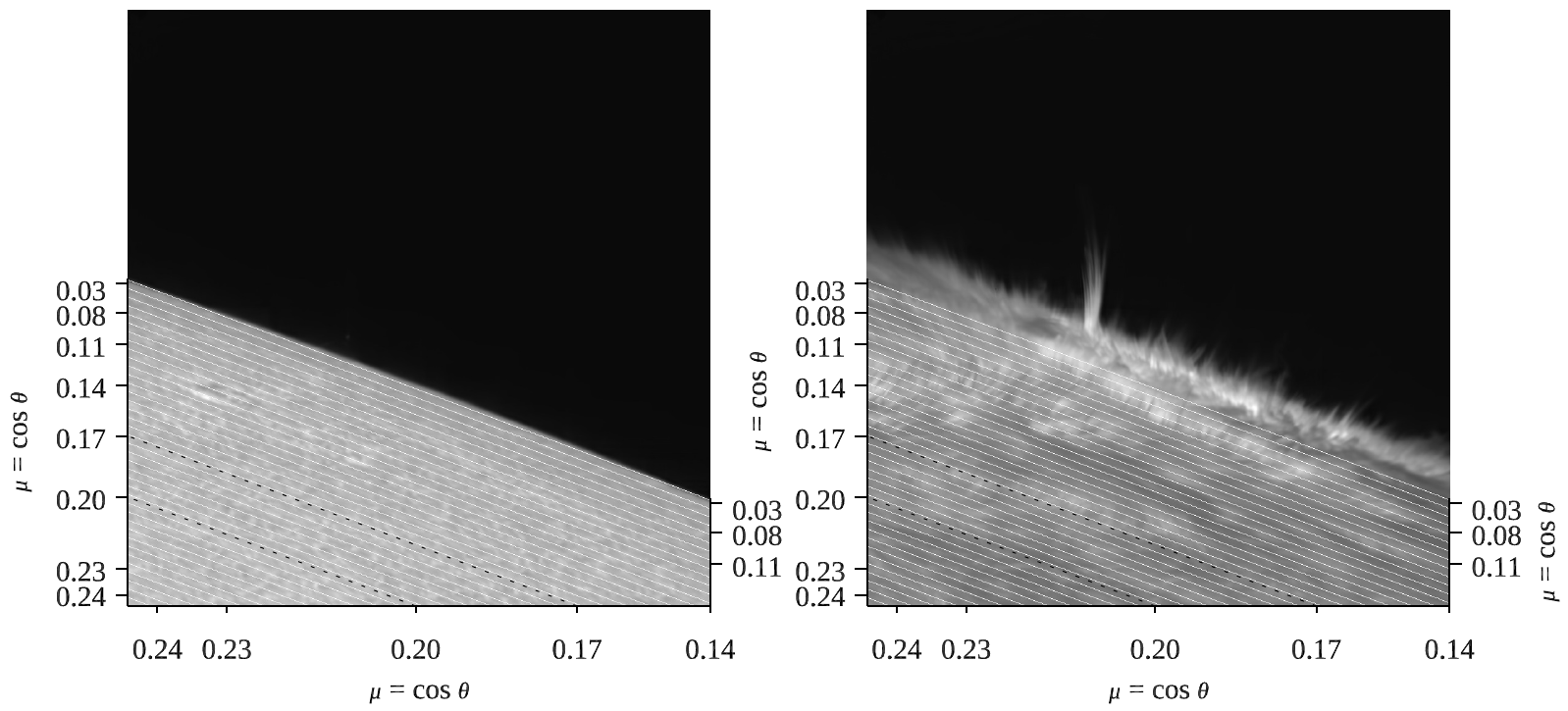}
\includegraphics[bb = 20 0 240 164, width=0.370\textwidth]{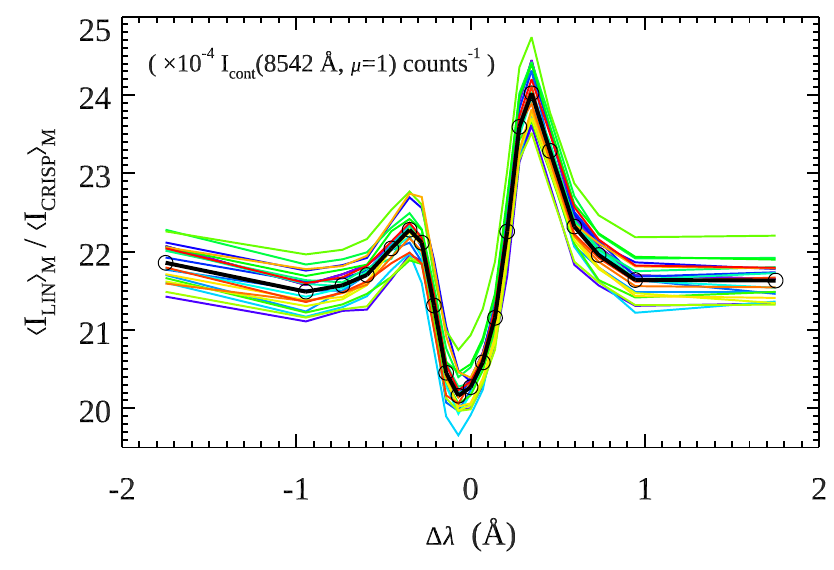}
\caption{Radiometric calibration of SST/CRISP \cawav\ data. Composite
  of \ca\ wing images at $\Delta\lambda = \pm 1.75$\,\AA\ (left) and
  line core image (middle) at 08:32:38\,UT. The mesh of white lines
  marks a sample of pixel lines labeled with the direction cosines
  $\mu = \cos\theta$ and used for computing radial intensity
  gradient. The black dotted lines define the area and the range of
  $\mu \in (0.17, 0.20) \equiv M$ adopted for computing the
  calibration profiles in the right panel. Right: Calibration
  profiles. Rainbow colors illustrate times of individual SST/CRISP
  scans. Blue, yellow, and red correspond to start, middle, and end of
  observations. The black curve is an average of the profiles. The
  black circles indicate the wavelength sampling of the observed
  \ca\ profile.}
\label{fig2}
\end{figure}

There is a wide variety of spicular jets in the solar atmosphere
observed in strong chromospheric lines. Their classification is based
on their morphological properties. However, there is no final
consensus regarding their categorization. A traditional spicule is
defined as a chromospheric jet with height between 6\,500 and
9\,500\,km, width between $700 - 2\,500$\,km and lifetime of up to $10
- 15$\,minutes \citep{Beckers1972,Pasachoff2009}. They are
characterized by rising and falling motions with speeds of $\sim 20 -
40$\,\kms. Disk observations have revealed similar types of
chromospheric fine structures such as quiet Sun mottles
\citep{Suematsu1995} and active region dynamic fibrils
\citep{dePontieuetal2004,Hansteenetal2006}. At the limb
\citet{dePontieu2007} discovered energetic and short-lived spicular
features called type II spicules. Soon after, structures with similar
properties were identified in high resolution ground-based
observations on the solar disk. These absorption features detected in
the blue and red wings of the \ca\ and \ha\ lines are called rapid
red-shifted and blue-shifted excursions
\citep{Langangen2008,Voort2009,Kuridze2015}.

Spicular structures are often observed with inverted Y-shaped
footpoints both at the limb and on the disk
\citep{YokoyamaShibata1995,Shibata2007,Heetal2009,Morita2010,Nishizuka2011,Nelsonetal2019}. Due
to their similarity with coronal X-ray jets,
\citet{YokoyamaShibata1995} and \citet{Shibata2007} refer to them as
anemone jets. The topology of these jets suggests that they are formed
as a result of magnetic reconnection in the lower solar atmosphere.

A reliable quantitative measurement of the physical parameters in
spicules is extremely challenging. The difficulties arise due to the
fact that they are very dynamic features, with spatial and
  temporal scales that are often close to the resolution limit of
modern solar telescopes and the large number of spicules in the solar
atmosphere makes overlapping emissions coming from them difficult to
separate. Chromospheric spectral lines are sensitive to plasma
properties in spicules. Therefore, chromospheric spectroscopy offers
the most powerful diagnostic opportunities. However, modeling and
interpretation of chromospheric spectral lines are always challenging
as they require solving the complex non-LTE (i.e., departures from
Local Thermodynamic Equilibrium) radiative transfer problem.

It is widely accepted that the external radiation field irradiating
spicules is entirely photospheric \citep{Beckers1968} and their
emission in most chromospheric lines forms under strong non-LTE
conditions. The line profiles of spicules in the strong chromospheric
spectral lines of hydrogen Balmer series, single ionized calcium
lines, and helium lines appear in emission 
with central reversal and broad, enhanced, asymmetric wings very
often \citep{Becketal2016}. As discussed by \citet{Beckers1972} and
confirmed by subsequent studies, the central reversals in spicule
spectra are not produced by the reduction of the source function at
the line core (self-absorption), but rather related to their
non-thermal dynamics.

One of the first spectroscopic diagnostics of spicule parameters was
carried out by \citet{Beckers1968}. He assumed that the spicule
  is a cylinder with the diameter of 815\,km situated
vertically on the Sun and irradiated by solar
radiation. Theoretical relations between electron temperature $T_{\rm e}$ 
and electron densities $n_{\rm e}$ were calculated for the
intensities of \ha, \ca, and He spectral lines through solving the
radiative transfer equation and the statistical equilibrium equation
in non-LTE. Then, for the observed intensities at different heights
along spicules, $T_{\rm e} - n_{\rm e}$ curves were obtained and
interceptions between these curves for different lines were used for
diagnostics of temperature and electron density. The temperature of
the spicules reported by \citet{Beckers1968} between the heights of $2
- 8$\,Mm is $\sim 9\,000 - 16\,000$\,K and the electron densities
between $4 - 8$\,Mm are $1.5\times10^{11} -
4.3\times10^{10}$\,\cmcub. \citet{Alissandrakis1973} and
\citet{Kralletal1976} implemented the same method and obtained similar
results for electron density and temperature as
\citet{Beckers1968}. \citet{SocasNavarroAndElmore2005} used multiline
full Stokes observations of spicules in the \ca\,8498, 8542\,\AA, and
\heir\ lines to derive spicule properties. They found that the
\ca\ and \heir\ lines have almost identical widths suggesting that
most of the broadening in spicules are non-thermal, which in turn
indicates that the electron temperature of the spicule should be lower
than 13\,000\,K. Measurements of magnetic fields in spicules
  using spectropolarimetric single-line observations have been done in
  the \heir\ line
  \citep{TrujilloBuenoetal2005,Centenoetal2010,OrozcoSuarezetal2015},
  in the \cawav\ line \citep{Kriginsky2020}, and in the \he\ line
  \citep{LopezAristeAndCasini2005,Ramellietal2005,Ramellietal2006}. For
  more details on spicule spectropolarimetry see
  \citet{TrujilloBueno2010}.

To interpret the spectra of spicular structures observed against the
solar disk and appearing as absorbing features in chromospheric lines,
\citet{Beckers1964} developed a different method known as the ''cloud
model". The cloud model is a simple inversion technique that fits the
observed intensity contrast profile with four free parameters ---
the source function, the optical depth, the Doppler width, and the
Doppler shift. With these parameters one can determine several other
physical parameters of the structure. This method has been
successfully applied over the years to chromospheric structures
observed in
\ha\ \citep{Alissandrakis1990,Tsiropoula1997,Tziotziou2003}.

Since the launch of the \emph{Interface Region Imaging Spectrograph}
\citep[\emph{IRIS};][]{DePontieu2014}, spicules have been intensively
observed and investigated in several UV lines by \citet{Pereira2014}
and \citet{Skogsurd2015}. \citet{Alissandrikisetal2018} compared the
\ion{Mg}{2}~k and h line profiles observed with \emph{IRIS} with
computations from a 1D non-LTE model and estimated the
temperature and electron density from the lower to the upper part of
spicules to be between $\sim 8\,000 - 20\,000$\,K and
$1.1\times10^{11} - 4\times10^{10}$\,\cmcub, respectively. More
recently, \citet{Tei2020} used 1D non-LTE vertical
slab models in single- and multiple slab configurations
and concluded that the width of the \ion{Mg}{2}~k and h line profiles
can be significantly influenced by a superposition of multiple
spicules along the LoS.

An alternative method of plasma diagnostics in spicules is the
so-called magnetoseismology, which relies on observations of
magnetohydrodynamic waves to infer the properties of a magnetic flux
tube \citep{Zaqarashvili2007,Zaqarashvili2009,Verth2011,Kuridze2013,Morton2014}. These methods also depend on the
assumed nature of the wave modes (local tube modes versus genuine
Alfv\'{e}n waves in more homogeneous media).

Spicule temperature and density were also determined using radio data
obtained with the Atacama Large Millimeter/submillimeter Array
\citep[ALMA;][]{WoottenThompson2009} radio telescope by
\citet{Shimojo2020}.  They derived kinetic temperature and the number
density of ionized hydrogen in the spicule plasma seen in the ALMA
100\,GHz images as 6\,800\,K and $2.2\times10^{10}$\,\cmcub.

The vertical stratification of the physical parameters along
chromospheric structures can be obtained using semi-empirical models
that attempt to reproduce the observed profiles in non-LTE radiative
transfer. A powerful way to construct atmospheric models with a
semi-empirical approach is to fit the observed Stokes profiles using
inversion algorithms \citep{delaCruz2017}. Through such inversions,
the ionization equilibrium, statistical equilibrium, and radiative
transfer equations are solved numerically to synthesize the Stokes
profiles under a set of predefined initial atmospheric conditions. The
\ca\ infrared (IR) triplet line at 8542\,\AA\ is well suited for the
development of such models due to its sensitivity to physical
parameters in the solar photosphere and chromosphere
\citep{Pietarila2007,Cauzzi2008}. Furthermore, calcium is singly
ionized under typical chromospheric conditions, with negligible
partial redistribution effects for the \ca\ IR lines
\citep{Uitenbroek1989}, making their modeling and interpretation of
the observations more straightforward. The non-LTE radiative transfer
code NICOLE \citep{SocasNavarro2000, SocasNavarro2015} is a suitable tool for our purposes
  here since it allows for the synthesis and inversion of \ca\ Stokes
  profiles and the construction of semi-empirical
models. Inversions with NICOLE have been successfully applied to
  the study of spectropolarimetric \cawav\ observations of umbral
flashes in sunspots \citep{delaCruz2013}, granular-sized magnetic
elements (magnetic bubbles) in an active region \citep{delaCruz2015},
and chromospheric flares \citep{Kuridze2017,Kuridze2018}. However,
NICOLE was not designed to work on off-limb spectroscopic
observations as it assumes a geometry in which the emergent
  radiation comes from the bottom of the atmosphere and hydrostatic
  equilibrium is imposed at each inversion iteration to compute the
  plasma density and pressure.

In this paper we use a new version of NICOLE to invert high-resolution
imaging spectroscopy in the \cawav\ line to construct models of the
solar off-limb spicule (Figure~\ref{fig1}). This new version
  first computes the atomic level populations in a vertically
  stratified atmosphere, including the spicule up to 10~Mm. Then, once
  all the source functions and opacities are known, a final formal
  solution computes the spectra in the direction of the line of sight,
  which in this case is horizontal, at a discrete set of heights
  specified by the user (in our case, the 14 points where observed
  spectra are available). These 14 profiles are fitted simultaneously
  with one single model atmosphere. The model has two components, as
  explained below. The chosen parameterization allows us to retrieve the height stratification of 
  temperature, mass and electron density, LoS velocity, and optical thickness along the spicule.

\begin{figure}
\includegraphics[width=\textwidth]{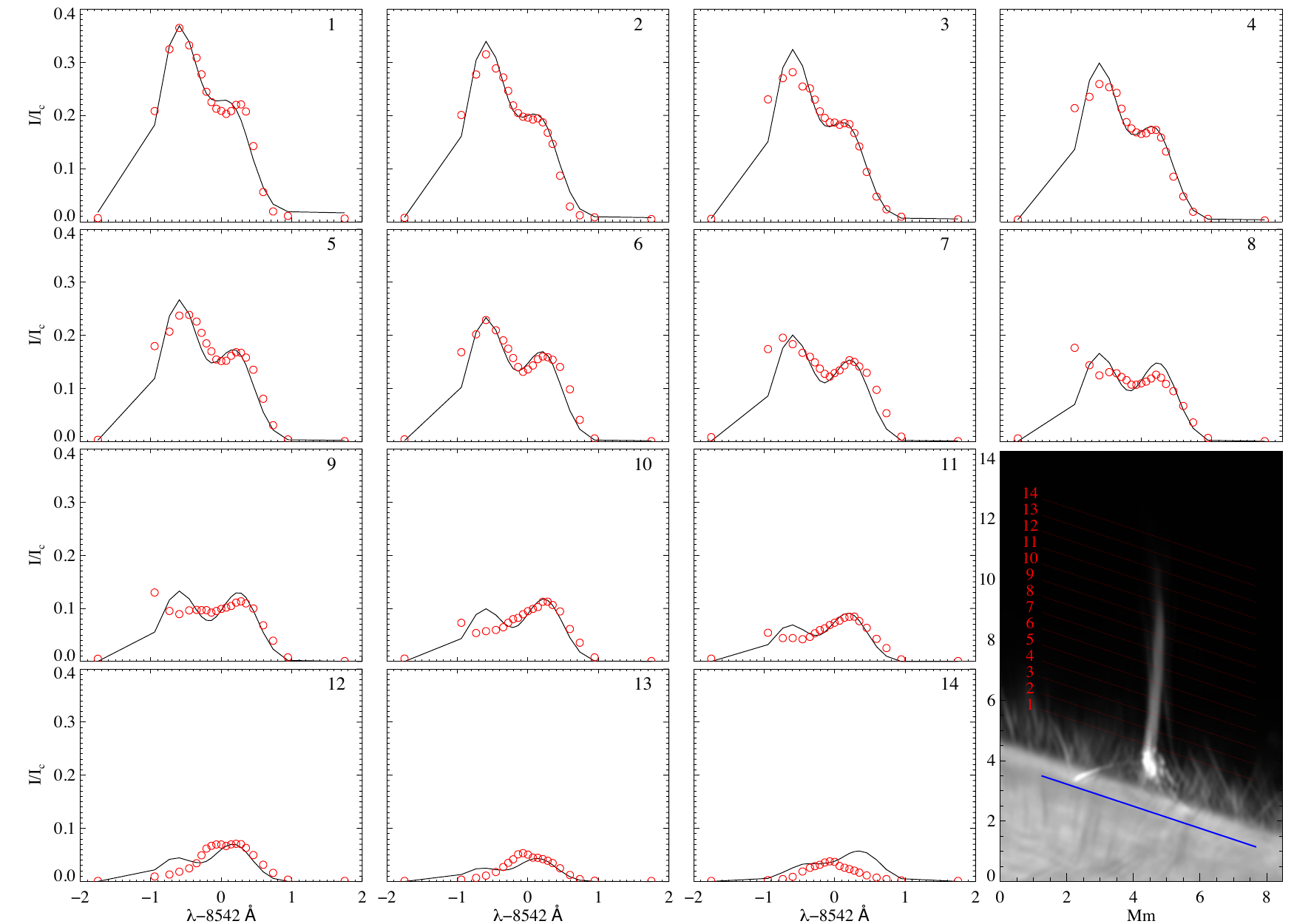}
\caption{Observed (red circles) and best-fit synthetic (solid black
  lines) \cawav\ line profiles of the spicule at fourteen different
  positions along its axis at 08:34:27\,UT. The numbers from 1
  (spicule footpoint) to 14 (spicule top) in the top right corners
  indicate the heights at which the line profiles along the
    spicule (bottom right panel) are located. The intensity of
  line profiles is normalized to the quiet Sun continuum intensity
  $I_{\rm c}$ at 8542\,\AA\ in the disk center. Bottom right:
    The spicule in the blue wing of \cawav\ at $\Delta\lambda =
    -0.735$\,\AA. The red lines identify the heights of origin of the
    spicule profiles $1-14$. The blue line shows the location of the
    photosphere at $\tau_{\rm 5000} = 1$ at $\mu=1$. The tickmarks of
    $y$ axis are labeled in Mm.}
\label{fig3_1}
\end{figure}

\begin{figure}
\includegraphics[width=\textwidth]{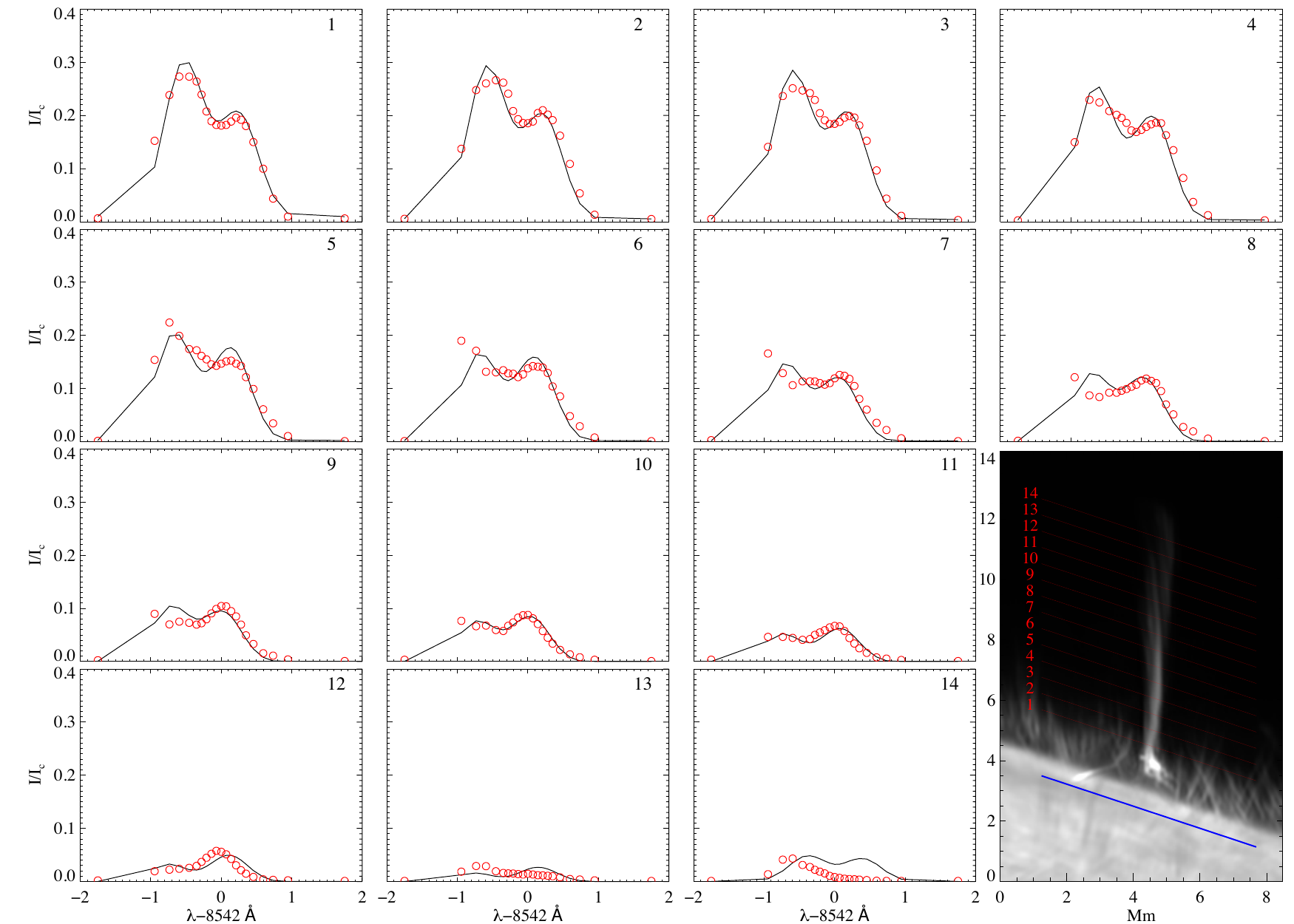}
\caption{Same as Figure~\ref{fig3_1} but at 08:35:33\,UT.}
\label{fig3_2}
\end{figure}

\begin{figure}
\includegraphics[width=\textwidth]{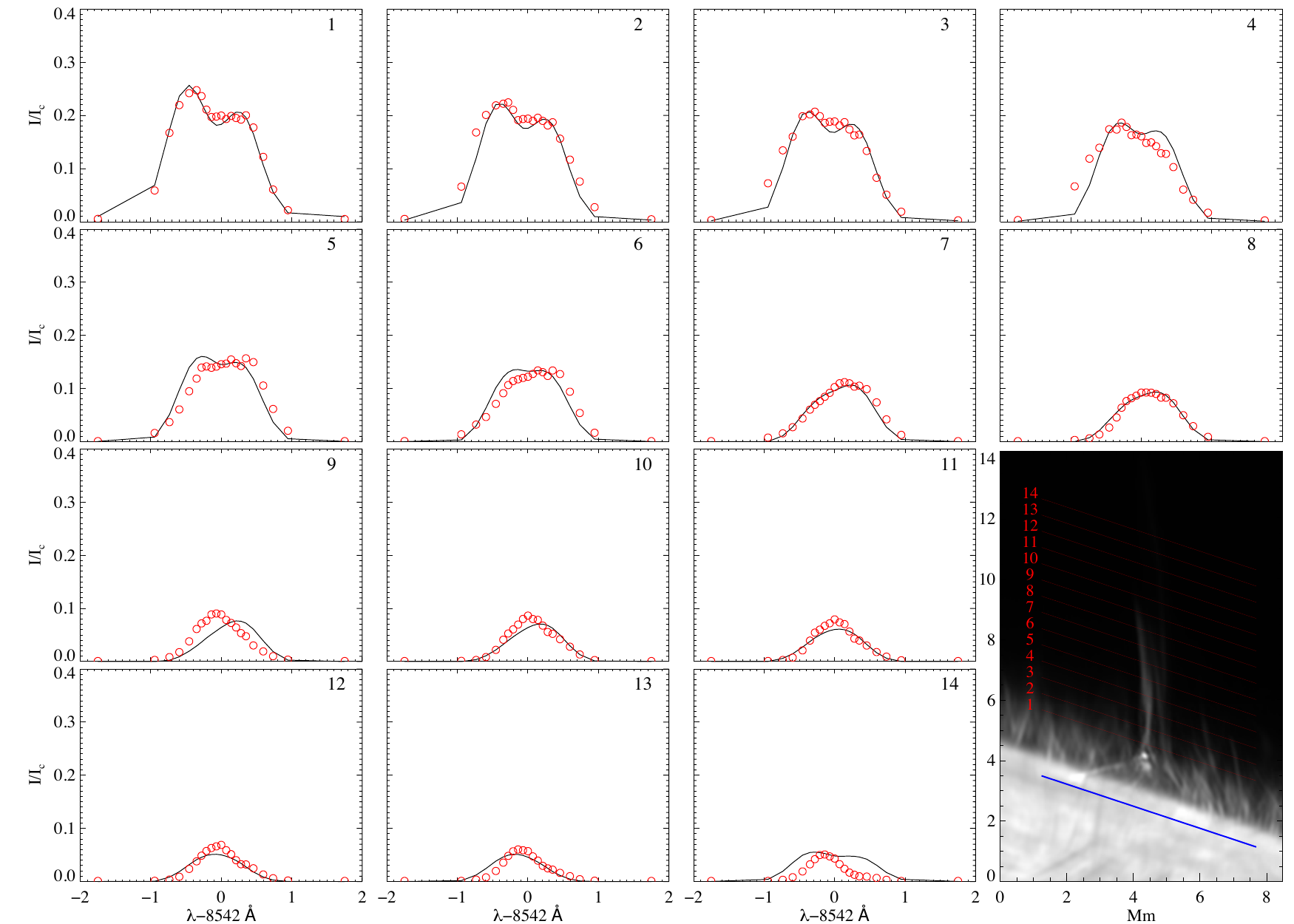}
\caption{Same as Figure~\ref{fig3_1} but at 08:38:18\,UT.}
\label{fig3_3}
\end{figure}

\begin{figure}
\includegraphics[width=\textwidth]{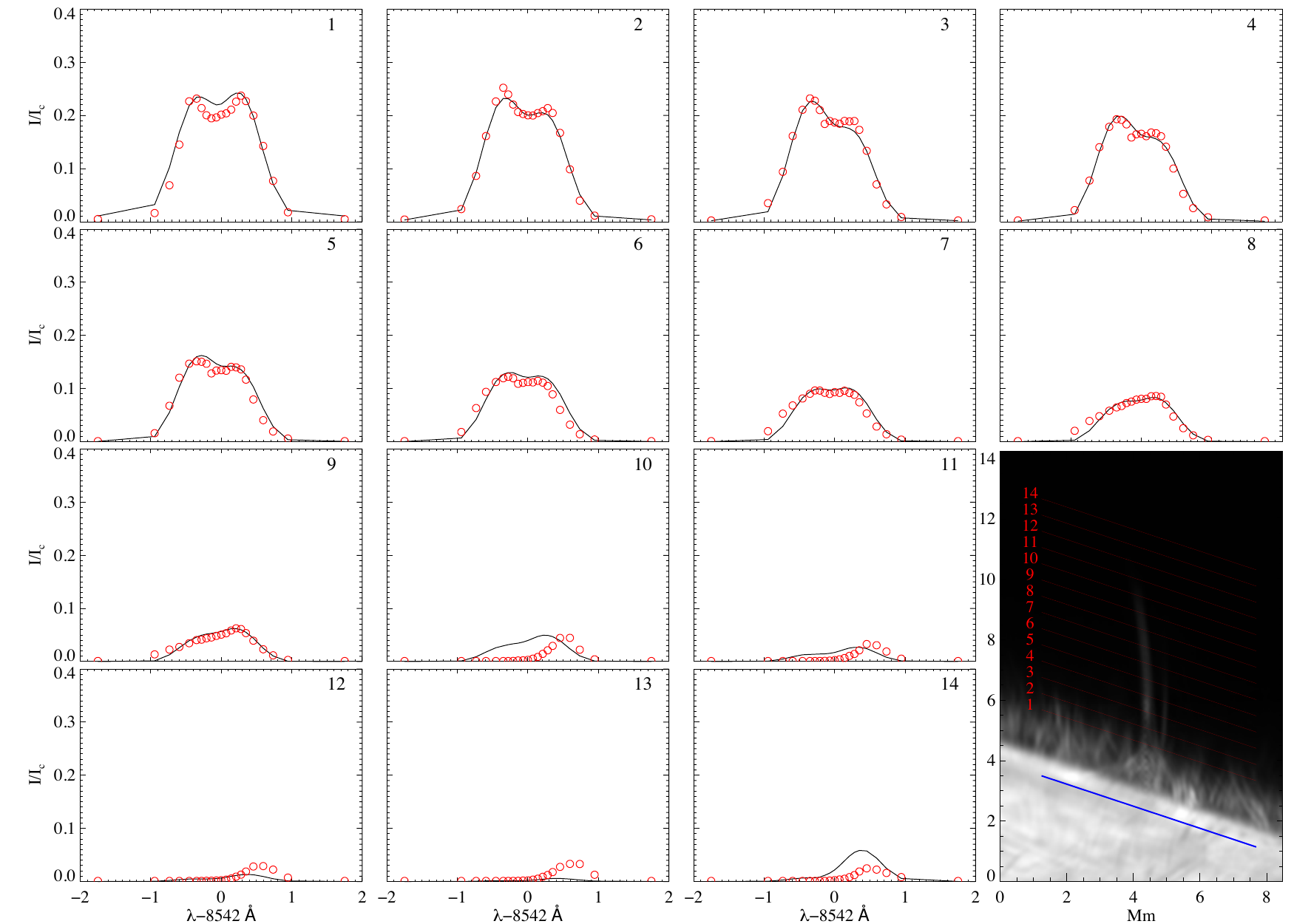}
\caption{Same as Figure~\ref{fig3_1} but at 08:41:03\,UT.}
\label{fig3_4}
\end{figure}


\section{Observations and Data Reduction}

\subsection{Observational setup}

We observed the west limb of the Sun between 08:20 and 08:43\,UT on
2018 June 22 near the active region (AR) NOAA 12714. The
heliocentric coordinates of the center of the field of view (FoV) at
the beginning of the observations were [937\arcsec,
  127\arcsec]. Observations were made with the CRisp Imaging
SpectroPolarimeter \citep[CRISP;][]{Scharmer2006,Scharmeretal2008} and
the CHROMospheric Imaging Spectrometer (CHROMIS) instruments, both
based on dual Fabry-P\'{e}rot interferometers (FPI) mounted on SST.
The imaging setup includes a dichroic beamsplitter with the
transmission/reflection edge at 5000\,\AA. CRISP is mounted in the
reflected red beam and CHROMIS in the transmitted blue beam
\citep[Figure~2]{Lofdahletal2018}.

The CRISP data comprises narrow-band imaging spectropolarimetry in the
\cawav\ line profile sampled from $-1.75$\,\AA\ to $+1.75$\,\AA\ in 21
line positions $\pm 1.75$, $\pm 0.945$, $\pm 0.735$, $\pm 0.595$, $\pm
0.455$, $\pm 0.35$, $\pm 0.28$, $\pm 0.21$, $\pm 0.14$, $\pm 0.07$,
0.0\,\AA\ from the line center (hereafter, unless specified
otherwise, when referring to the \ca\ line we mean the \cawav\ line).
Each spectral scan of the \ca\ line had an acquisition time of 16\,s
but the cadence of the CRISP time series was 33\,s due to the inclusion of
spectropolarimetric scans in the \ion{Fe}{1}~6302\,\AA\ photospheric
line. However, we note that the present paper includes only the
analysis of the \ca\ data, as the spicule analyzed in this study and
its footpoints were not detected in the \ion{Fe}{1} line. The CRISP
data are processed by the CRISPRED reduction pipeline
\citep{delaCruzRodriguezetal2015} and reconstructed with Multi-Object
Multi-Frame Blind Deconvolution
\citep[MOMFBD;][]{Lofdahl2002,vanNoortetal2005}.

Simultaneous observations were taken with the CHROMIS imaging
spectrometer -- a dual FPI observing in the range of $3\,900 -
4\,900$\,\AA.  The CHROMIS observations comprise narrow-band and
wide-band spectral imaging in the \hb\ and
\ion{Ca}{2}\,H~3968.5\,\AA\ lines, plus one position in the continuum
at 4000\,\AA. CHROMIS data are processed using the CHROMISRED
reduction pipeline, which includes MOMFBD image restoration
\citep{Lofdahletal2018}. 

In this work we focus only on the inversion of the CRISP \ca\ Stokes
{\it I} profiles.

\subsection{Data Radiometric Calibration} 

The inversion of SST/CRISP spectroscopic observations of the
spicule requires careful intensity calibration of the
data. Inversion codes, including NICOLE, usually require the
input data to be calibrated with respect to the quiet-Sun
continuum intensity at the disk center to convert the digital
  counts in the observations to actual physical units. However, this
  was not an option in our case due to the absence of a disk-center
  continuum reference in the data.

To calibrate a single CRISP \ca\ scan, the Stokes
{\it I} wing images at $\Delta\lambda = \pm 1.75$\,\AA\ are averaged
yielding a composite image (Figure~\ref{fig2}: left panel). The limb
is identified in the composite image by the IDL function
\texttt{sobel.pro} and approximated by a limb line. Then a dense mesh
of 536 lines parallel to the limb is created covering the whole
visible disk. Using the solar disk radius of 944\farcs338 and the
CRISP spatial sampling of 0\farcs057\,pixel$^{-1}$ the position
cosines $\mu = \cos\theta$ corresponding to the mesh lines are
calculated. A sample of 38 equidistant mesh lines and the position
cosines are shown in Figure~\ref{fig2} (left and middle
panel). Averaging Stokes {\it I} pixel intensities over particular
mesh lines at a given $\mu$ yields radial intensity gradients $I_{\rm
  CRISP}(\lambda_{\rm CRISP}, \mu)$ at particular CRISP wavelengths,
$\lambda_{\rm CRISP}$. Quasi-linear intensity segments $I_{\rm
  CRISP}(\lambda_{\rm CRISP}, M)$ (not shown here) in the position
cosines ranging from 0.17 to 0.20 (hereafter $M \equiv \mu \in (0.17,
0.20)$) are adopted for calibration, avoiding the remnant of
the AR NOAA 12714 seen at $\mu \sim 0.13$ (Figure~\ref{fig1}: bottom
right panel and Figure~\ref{fig2}: left panel). The calibrated
\ca\ profiles from \citet{Linskyetal1970} are adopted as a reference
extrapolating them for $M$ and convolving them with the transmission
profile of the CRISP Fabry-P\'{e}rot etalons provided by J.~de la Cruz
Rodr\'{\i}guez (2017, private communication). The resulting profiles
are referred to hereafter as $I_{\rm LIN}(\lambda_{\rm LIN},
M)$. To compensate for differences between the CRISP wavelength scale
$\lambda_{\rm CRISP}$ and the scale $\lambda_{\rm LIN}$ used in
\citet{Linskyetal1970}, the precise line minimum positions
$\lambda_{\rm c}$ of $I_{\rm CRISP}(\lambda_{\rm CRISP}, \mu)$ are
measured in a broad range of $\mu$ values. Then the profile minimum
positions of $I_{\rm LIN}(\lambda_{\rm LIN}, M)$ are compared
to $\lambda_{\rm c}$ at the corresponding $\mu$, allowing the
  interpolation of $I_{\rm LIN}(\lambda_{\rm LIN}, M)$ on the CRISP
wavelength scale. This yields the reference intensity gradients
$I_{\rm LIN}(\lambda_{\rm CRISP}, M)$. The ratio of the reference
intensity gradient $\langle I_{\rm LIN}(\lambda_{\rm CRISP}, M)
\rangle_M$ and the observed gradient $\langle I_{\rm
  CRISP}(\lambda_{\rm CRISP}, M) \rangle_M$, averaged over the $M$
range, renders a calibration profile (Figure~\ref{fig2}: right panel)
allowing the conversion of the \ca\ Stokes {\it I} intensities from
digital numbers to the physical units expressed relative to the quiet
Sun continuum intensity at 8542\,\AA\ in the disk center $I_{\rm c}$
(Figures~\ref{fig3_1} -- \ref{fig3_4}). To compensate for possible
variations in sky transparency, the calibration procedure is repeated
separately for each CRISP scan. The right panel of Figure~\ref{fig2}
shows a cluster of profiles used for calibrating particular scans.

\subsection{Corrections for canopy emission and stray light}
\label{stray}

We subtracted two non-spicular components from the spicule
  spectra -- average foreground canopy emission and off-limb stray
  light. The height of the chromospheric canopy, populated by short
  type II spicules and the background
  chromosphere, was estimated as $\sim 2\,800$\,km above the
  photospheric limb by the \ca\ line center images. Then the reference
  canopy \ca\ profiles $I_{\rm ref}(h, \lambda)$ were created for each
  height $h$ between the photospheric limb and the top of the
  canopy. Assuming that only foreground canopy emission contributes to
  the spicule emission, the intensity $0.5I_{\rm ref}(h, \lambda)$ was
  subtracted from the spicule profile at the same height between the
  photospheric limb and the canopy top. Above the canopy, the stray
  light intensity $I_{\rm stray}(h, \lambda)$ was determined by
  intensity gradients at particular heights and wavelengths and
  subtracted from the spicule profiles at the same heights. We note
  that the canopy and stray-light emissions are significantly weaker
  compared to the emission that comes from the 
  spicule due to its exceptional brightness and clear
  separation. Therefore, the subtractions modify the spicule
    spectra only slightly.

%
\section{Spicule appearance and line profiles}

Figure~\ref{fig1} shows images of the 
spicule in the \ca\ and \hb\ line core and blue wing images
taken at 08:34\,UT. The spicule was already developed at the
start of the observations at $\sim$\,08:20\,UT and disappeared after
around 20\,minutes. The line core images show that the structure
extends up to the height $\sim9$\,Mm above the limb, well above the
chromospheric canopy populated with shorter type II spicules
(Figure~\ref{fig1}). The blue wing images nicely illustrate the
  spicule footpoint shaped as the inverted Y. It suggests an
  association of the spicule with small-scale magnetic reconnection,
  which likely launched the spicule. Thus the event discussed here is
  of the same type and origin as those in \citet{Shibata2007},
  \citet{Heetal2009}, \citet{Nishizuka2011}, and
  \citet{Nelsonetal2019}. The spicule has an inclination angle
$\theta\sim70\degr$ with respect to the limb.  
The average width of the spicule is around 1\,Mm.

We select four line profile scans taken at $\sim$\,08:34, 08:35,
08:38, and 08:41\,UT when the spicule was
relatively bright and started becoming fainter. The red circles in
Figures~\ref{fig3_1} -- \ref{fig3_4} represent the observed \ca\ line
intensities of the spicule for the four selected snapshots at fourteen
space-separated positions along its axis.  They are normalized with
respect to the quiet-Sun continuum intensity at 8542\,\AA\ in the disk
center $I_{\rm cont}(8542\,\text{\AA}, \mu=1) \equiv I_{\rm c} $.
The red lines in the lower right panels of Figures~\ref{fig3_1}
  -- \ref{fig3_4} show the equidistant heights located at different
  positions along the spicule.  Along each height the pixel with
  maximum wavelength integrated intensity are found and line profiles
  for those pixels are selected for inversions.  The lowest (1) and
highest (14) positions (the lower right panels in Figures~\ref{fig3_1}
-- \ref{fig3_4}) are located at the heights $\sim2\,050$ and
8\,600\,km above the limb, respectively. It is known that the
  limb is uplifted by about $\sim 350$\,km above the base of the
  photosphere at $\tau_{\rm 5000} = 1$ measured at $\mu=1$
  \citep[Table~1]{Lites1983}. This off-set is due to an increase of
  optical depth at 5000\,\AA\ when looking from disk center ($\mu=1$)
  to limb ($\mu=0$).  The edge of the solar disk at the
  \ca\ $\pm1.75$\,\AA\ line positions is also expected to be formed
  at $\sim 350$\,km above the $\tau_{\rm 5000} = 1$ at $\mu=1$. The
  blue lines in the lower right panel of
  Figures~\ref{fig3_1} -- \ref{fig3_4} show the location
  of the true photosphere and the indicated
  heights of the selected pixels along the spicule are considered as a
  projection on the normal to the photospheric limb.

Figures~\ref{fig3_1} -- \ref{fig3_4} show that the line profiles
appear in emission with a central reversal and broad, enhanced,
asymmetric wings. However, the observed central reversals are not due
to self-absorption by spicule plasma itself.  The main characteristic
of the self-absorption is that blue-shifted line core (the location of
the central dip) produces a red asymmetry (higher red wing emission)
and the red-shifted line core produces a blue asymmetry (higher blue
wing emission). This is a well known phenomenon related to
the shift in the wavelength of maximum opacity in the
line core to shorter and longer wavelengths, which invokes the red and
blue asymmetries, respectively \citep{Kuridze2015b}.  The observed
line profiles do not show this behavior suggesting that the central
reversal could be due to strong opposite LoS velocity components
within the spicule. 
The top part
     of the spicule at 08:38 and 08:41\,UT has single peak line
     profiles without central reversal (Figures~\ref{fig3_3}
     and \ref{fig3_4}).  Intensities at all wavelength
     positions across the \ca\ line decrease as a function of height
     (Figures~\ref{fig3_1} -- \ref{fig3_4}).  The maximum line
     emission in the blue wing at around
     $-0.7$\,\AA\ from the line center drops by about a factor
     of $\sim3 - 10$ over the heights between $\sim2 - 9$\,Mm above
     the limb (Figures~\ref{fig3_1} -- \ref{fig3_4}).

\section{NICOLE inversion code and spicule model}

\subsection{Modified version of NICOLE }
\label{nic}

NICOLE \citep{SocasNavarro2015} solves the multilevel non-LTE
radiative transfer problem to invert or synthesize photospheric and
chromospheric spectral lines \citep{SocasNavarro1997}.  In the
inversion mode the code iteratively perturbs physical parameters such
as temperature, LoS velocity, magnetic field strength, and
microturbulence in a predefined set of points (nodes) of an initial
guess model atmosphere to find the best match with the observations
\citep{SocasNavarro2000}. We use here a five bound
level-plus-continuum model of single ionized calcium atom with
complete angle and frequency redistribution, which is applicable to
lines such as \cawav\ \citep{Uitenbroek1989}.

As discussed above, NICOLE was not designed originally for
  inversions of spectra observed off the limb. A new version has
  been developed for this work with the following alterations. 
  The model atmosphere may extend to heights of several Mm. 
  The lower layers of the model up to 2\,Mm 
  represent the underlying average atmosphere and provide the
  radiation field illuminating the spicule from below. This lower part
  of the model is treated as before, except that it is kept fixed in
  the inversion. We chose the well-known 
  Harvard-Smithsonian Reference Atmosphere model 
  \citep[HSRA;][]{Gingerich1971} as a good representation of the
  average solar atmosphere.

  The atmosphere above 2~Mm is identified with the spicule and will be the
  subject of our study. We initialize this upper part of the model
  with an extrapolation of HSRA. This model may be modified according
  to a number of free parameters that will be optimized by the code to
  provide the best fit to the observations in a least-squares
  sense. Motivated by the double-peak appearance of the profiles as
  well as some previous works in the literature, we decided to
  implement a two-component scenario. All the parameters that define
  the spicule model are duplicated (both components are, in principle,
  independent) and an additional parameter, the filling factor of each
  component, is introduced. This parameter is also
  height-dependent.

  Each inversion iteration requires one or more calculations of
  synthetic profiles. The synthesis proceeds in two stages. In the
  first stage, an iteration scheme computes the atomic level
  populations that are self-consistent with the radiation field that
  they produce in this 1D model. The 1D approximation implies that
  each grid point sees the layers above and below as being infinite in
  the horizontal direction. This is a good approximation if the layers
  are optically thick because in this case one is not able to see the
  edge of the spicule. As we shall see below, this assumption is
  reasonably well justified.

We do not assume hydrostatic equilibrium in the spicule. The mass
density is recovered 
as a smooth function
interpolating between a set of nodes equispaced vertically
along spicule axis. The number of nodes for the height-dependent variables is defined by the user.

The second stage computes the synthetic profiles with a final formal
  solution of the radiative transfer equation and convolves them with the CRISP transmission profile.
  In this case, the
  radiative transfer is solved in the horizontal direction. The
  boundary condition for this transfer is that there is zero radiation
  field incoming from behind the spicule. We simply have to integrate
  at each height of interest in the direction perpendicular to the
  axis between the back of the spicule and the front. For this reason,
  the geometrical width of the spicule is relevant. It is a free
  parameter of the model (one for each component).
The assumption of no incident radiation from behind the spicule is
justified because the main body of the observed spicule is located
above the chromospheric canopy and the forest of type~II
spicules. Therefore, there are no overlapping spicules along the
LoS. Furthermore, the emission from the background and foreground
corona and chromospheric canopy can be safely ignored as it has
$\sim20-200$ times lower emission at $\lambda = 8542$\,\AA\ than
the spicule itself in relevant heights, and these
  emissions were subtracted from the spicule spectra
  (Section~\ref{stray}).

\subsection{Double-component spicule model}
\label{par}

\begin{table}
  \centering
  \caption{Parameters of spicule model employed in the inversion.}
  \label{tab}
  \begin{tabular}{l   lc}
    \hline
    \hline  
    Free parameters \\
    \hline
    Temperature, $T$                  & 1 node \\
    Mass density\tablenotemark{\footnotesize a}, $\rho$  &  3  nodes  \\
    LoS velocity, $V_{\rm LoS,red,blue}$  &  3 nodes  \\
    Filling factor, $F_{\rm red,blue}$     &  3 nodes  \\
    \hline
    Fixed parameters \\      
    \hline
    Spicule thickness, $D$           &   1\,000\,km  \\
    Doppler width, $W_{\rm Dopp}$       &  150\,m\AA\  \\
    %
    %
    Underlying atmosphere            &  HSRA  \\
    \hline
  \end{tabular}
  \tablenotetext{a}{ -- Electron density $n_{\rm e}$, gas pressure $p_{\rm gas}$, and electron pressure $p_{\rm e}$ are computed from the mass density $\rho$ and the equation of state.} 
\end{table}

The free parameters that characterize the model are the
  following. A user-defined number of nodes defines the
  height-dependent parameters: mass density $\rho$, line-of-sight
  velocity $V_{\rm LoS}$ and filling factor of the first component.
  The height-independent parameters are: temperature $T$,
  spicule thickness $D$, Doppler width $W_{\rm Dopp}$.

Inspired by the double-peak appearance of \ca\ profiles
  (Figures~\ref{fig3_1} -- \ref{fig3_4}), we invoke a two-component spicule model, which
  represents separately the redshifted (red component) and blueshifted
  (blue component) lobes of spicule \ca\ profiles by corresponding
  sets of height-dependent ($\rho$, $V_{\rm LoS}$) and
  height-independent ($T$, $D$, $W_{\rm Dopp}$)
  parameters. With this scheme there are many free parameters. We
    sought to impose reasonable constrains on the model to reduce the
    degrees of freedom.  Following the result by \citet{Beckers1968},
  claiming that a model with constant temperature
 can reproduce the spicule intensities in chromospheric lines (see
  Section~\ref{disc}), we considered to start with the ansatz that
  spicules are isothermal features, even with the same
  temperature in both model components. 
  We interpret the asymmetry in the peak
  intensities as due to different filling factors of the model
  components within a resolution element to represent different plasma
  volumes moving towards and away from an observer. As the peak
  intensity asymmetry varies with the height (Figures~\ref{fig3_1} --
  \ref{fig3_4}), we adopt height-dependent filling factors.


\begin{figure}
\includegraphics[width=\textwidth]{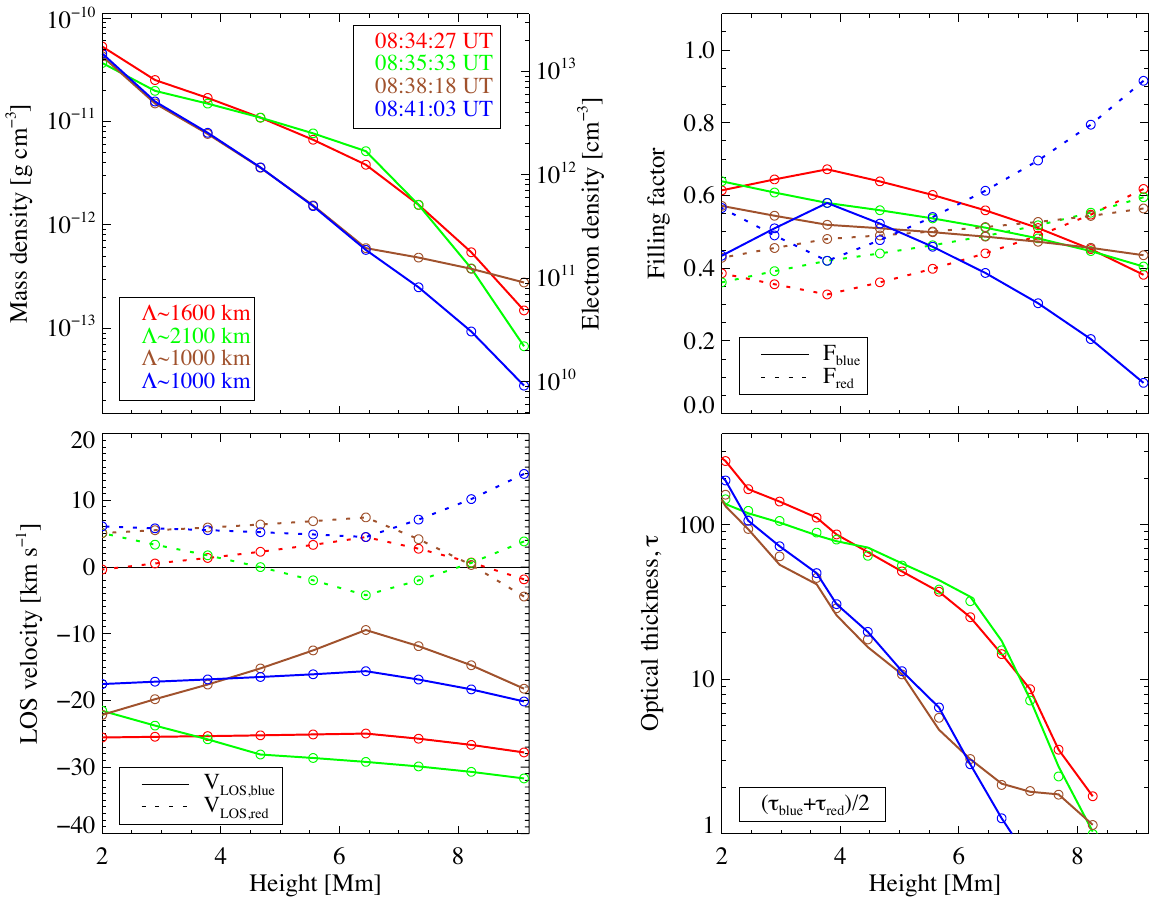}
\caption{Height stratifications of parameters of the
    double-component spicule model inferred by the NICOLE inversion
    for the times shown in the right corner of the top left panel. Top
    left: The mass and electron densities (right $y$
    axis). Corresponding height scales $\Lambda$ are shown in the
    bottom left corner. Bottom left and top right: The LoS velocities after compensating
    for the velocity of solar rotation
    and filling factors.  Solid and dashed lines correspond to the
    blue and red component of the spicule model, respectively. Bottom
    right: The average optical thickness $\tau$ in the \cawav\ line
    center as an average of the component optical thicknesses
    $\tau_{\rm blue}$ and $\tau_{\rm red}$.}
\label{fig6}
\end{figure}

In order to constrain the model further, we
impose that both components must have the same mass density
stratification. Three nodes are used to define its height dependence $\rho(z)$.
The stratifications of the electron density, electron and gas pressures
are computed from the equation of state using the mass density
obtained directly from inversions.
We adopt three nodes for the LoS velocity
of the blue model component $V_{\rm LoS,blue}$ and also three nodes
for the LoS velocity of the red component $V_{\rm LoS,red}$. It means
it can vary as a second-order polynomial with height.
We also assign three nodes for the height-dependent filling factors
$F_{\rm red}, F_{\rm blue}$ but constraining them by the condition
$F_{\rm red}+F_{\rm blue}=1$. Since the lower atmosphere (HSRA) is
unchanged and the code only modifies the spicule model, the nodes are
equispaced in the height scale $z$ starting from $z = 2$\,Mm.
Other parameters of the spicule model, such as spicule (i.e. segment)
thickness $D$ and Doppler width $W_{\rm Dopp}$ were fixed to the
values shown in Table~\ref{tab}, which summarizes the model parameters
and node numbers. Those values were found to be suitable choices after
some manual exploration of the parameter space.

\section{Inversion results} 

The new version of NICOLE fits simultaneously the observed
\ca\ profiles from fourteen different locations along the
spicule axis with the double-component model that has a constant temeprature and a
smooth height-dependent density.
The inversion is performed for four selected snapshots of the spicule
lifetime.  Figures~\ref{fig3_1} -- \ref{fig3_4} show the best-fit
synthetic profiles resulting from the inversion. All models in the
four selected snapshots characterize the constant temperature of $T =
9\,560$\,K in both components.

The height stratification of the mass density and the electron
  density, reproducing the fourteen observed \ca\ profiles at four
  selected snapshots, are shown in Figure~\ref{fig6} (top left
  panel). In order to compute a density scale height, we fitted
  an exponential function (not shown) to the density stratifications retrieved by
  the inversion. The density scale heights $\Lambda$ obtained in this manner
are shown in the bottom left corner. The height scale drops from about
2\,100\,km to about 1\,000\,km over the interval of 7\,minutes. The
inversion yields the mass density of components typically in the range
from about $4\times10^{-11}$\,g\,\cmcub\ at the spicule bottom to about
$3\times10^{-14}$\,g\,\cmcub\ at its top. The corresponding electron
densities computed from the mass densities by the equation of state
are shown on the right $y$ axis. They decrease from about
$1.4\times10^{13}$\,\cmcub\ at the spicule bottom to about
$9\times10^9$\,\cmcub\ at its top. The density drop in time is
apparent especially in the upper parts of the spicule.

The variations of the LoS velocity in the blue ($V_{\rm
    LoS,blue}$) and red ($V_{\rm LoS,red}$) components of the spicule
  model are shown in the bottom left panel of
  Figure~\ref{fig6}. Disregarding the height-time variations, we can
  take $-22$\,\kms\ as typical for the former and $+3$\,\kms\ for the
  latter. 
  We interpret the asymmetry in their absolute values as a
  consequence of vector superposition 
  of counter-oriented Doppler velocity components 
  of spicule tilted toward the observers. 
  The whole-spicule decrease of $|V_{\rm LoS,blue}|$
  from about 27\,\kms\ to 16\,\kms\ is followed by an increase of
  $|V_{\rm LoS,red}|$ from about 1\,\kms\ to 6\,\kms\ apparent below
  $\sim 6$\,Mm. This is compatible with the ceasing of upflow motion
  and also with the drop of density scale height $\Lambda$ over time.
  
The top right panel of Figure~\ref{fig6} shows height-time
  variations of the filling factor in the blue ($F_{\rm blue}$) and
  red ($F_{\rm red}$) component of the spicule model. It suggests
  roughly balanced $\sim 50/50$ coverage of the resolution element by
  the spicule model components at all heights. An exception is the
  last moment (08:41:03\,UT) shown in blue, when the red model
  component ($F_{\rm red}$) dominates over the resolution element at
  the heights above $\sim 7$\,Mm.

The bottom right panel of Figure~\ref{fig6} shows the average
  optical thickness of spicule $\tau$ in the \ca\ line center. Due to
  the same mass densities of both model components, the component
  optical thicknesses $\tau_{\rm blue}$ and $\tau_{\rm red}$ are also
  almost the same. Therefore the panel shows their simple average
  separately for the four selected times. The optical thickness
  decreases exponentially from about 300 at the spicule bottom to
  about one at its top. It proves that the spicule plasma is mostly
  optically thick in \ca\ and validates our assumption about
  infinitely extended plane-parallel spicule slab in the radiative
  transfer calculations (Section~\ref{nic}). The drop of $\tau$ over
  time in its upper part is also apparent.

We note that, the inversions can not achieve a good fit at the top part of the spicule (Figures~\ref{fig3_1} -- \ref{fig3_4}, points $13 - 14$). 
Due to the significant change in relative amplitudes of the two components at the top, 
the code can not fit the line profiles with the same success using the same parametrisation. 
The ratio of
amplitudes along the spicule is nearly constant but at the top one component vanishes and the ratio raises significantly. 
Possible reason could be the end of the helical structure
or spinning/unwinding motion, when the field does not 
wrap around any more at the top part of the spicule. 
We plan to explore this in more detail in the follow up paper where we use the 2D models.

\subsection{Response functions}

To investigate the sensitivity and response of the Stokes {\it
  I} \ca\ profiles to variations of physical parameters, we
investigate the numerical response functions (RFs) for the spicule
model implemented in the new version of NICOLE. The response function
$RF_x(\lambda)$ to the given atmospheric parameter $x$ is a measure of
how the line intensity at a given wavelength $\lambda$ reacts to a
small perturbation on $x$. Formally, $RF_x(\lambda)$ to the
parameter $x$ is defined as:
\begin{equation} RF_x(\lambda)=[I_{\rm pert}(\lambda)-I_{\rm ref}(\lambda)]/(x_{\rm pert}-x_{\rm ref})\,, 
\end{equation}
where $x_{\rm ref}$ is a reference value of a spicule model parameter
producing reference intensity $I_{\rm ref}(\lambda)$ at $\lambda$ and
$x_{\rm pert}$ is the value of perturbed parameter producing the
perturbed intensity $I_{\rm pert}(\lambda)$. In the limit 
when $x_{\rm pert}$ tends to $x$, the expression becomes the
  mathematical representation of a derivative. Figure~\ref{fig7}
shows the Stokes {\it I} RFs of the \ca\ line to perturbations of the
temperature, spicule thickness, and mass density for the fourteen
segments of the blue component of the spicule model. The curves are
normalized to maximum values of corresponding $RF_x$. All displayed
RFs show maxima or minima at $\Delta\lambda\sim -0.6$\,\AA. While the
RFs suggest different response of \ca\ blue wing intensity to
parameter perturbations in the bottom spicule segments $1-6$ there is
a strong degeneracy of response in most of the upper segments $6-14$
--- the blue wing intensity reacts the same to parameter perturbations
of the blue component model of spicule. This is the reason why
  it would not be possible to invert each spectrum separately. One
  needs the constraints given by a consistent height-dependent model
  to fit all the observations at various heights.

%
\section{Discussion and conclusions}
\label{disc}

We study an off-limb solar spicule using high-resolution imaging
spectroscopy obtained in the \cawav\ line by the SST/CRISP
instrument. We use a new version of the non-LTE code NICOLE assuming
a double-component spicule model to infer the spicule
  temperature and height stratifications of density, and LoS velocity
along the spicule axis defined by pixel positions at the maxima
of line integrated intensities.

Our analyses shows that the spicule plasma is isothermal having the
uniform temperature of 9\,560\,K along its length in the time interval
of 7\,minutes.  Early and also recent measurements reported
temperature variation along the spicule axis
\citep{Beckers1968,Beckers1972,Alissandrakis1973,Kralletal1976,MatsunoHirayama1988,Alissandrikisetal2018}.
However, despite being reported some temperature variations, 
based on analyses of the $T_{\rm e} - n_{\rm e}$ curves at
different heights, the pioneering work of \citet{Beckers1968}
concluded that the spicule intensities in chromospheric
lines do not yield information about temperature variation and hence
observed spectral characteristics in principle can be reproduced by a
spicule model with uniform temperature (see page 408 therein).  
In this context \citet{Kralletal1976} also mentioned that H$\alpha$ emission 
is a rather sensitive indicator of electron density and relatively independent 
of temperature over a large temperature range (see page 100 therein).  
Our results confirms this as we fitted multiple \ca\ profiles with
constant temperature in four selected snapshots from the
spicule lifetime. The inverted Y-shaped topology of the spicule
footpoint implies a magnetic reconnection in the junction point
that may have populated the spicule body with plasma of uniform temperature.

\begin{figure}
\includegraphics[width=\textwidth]{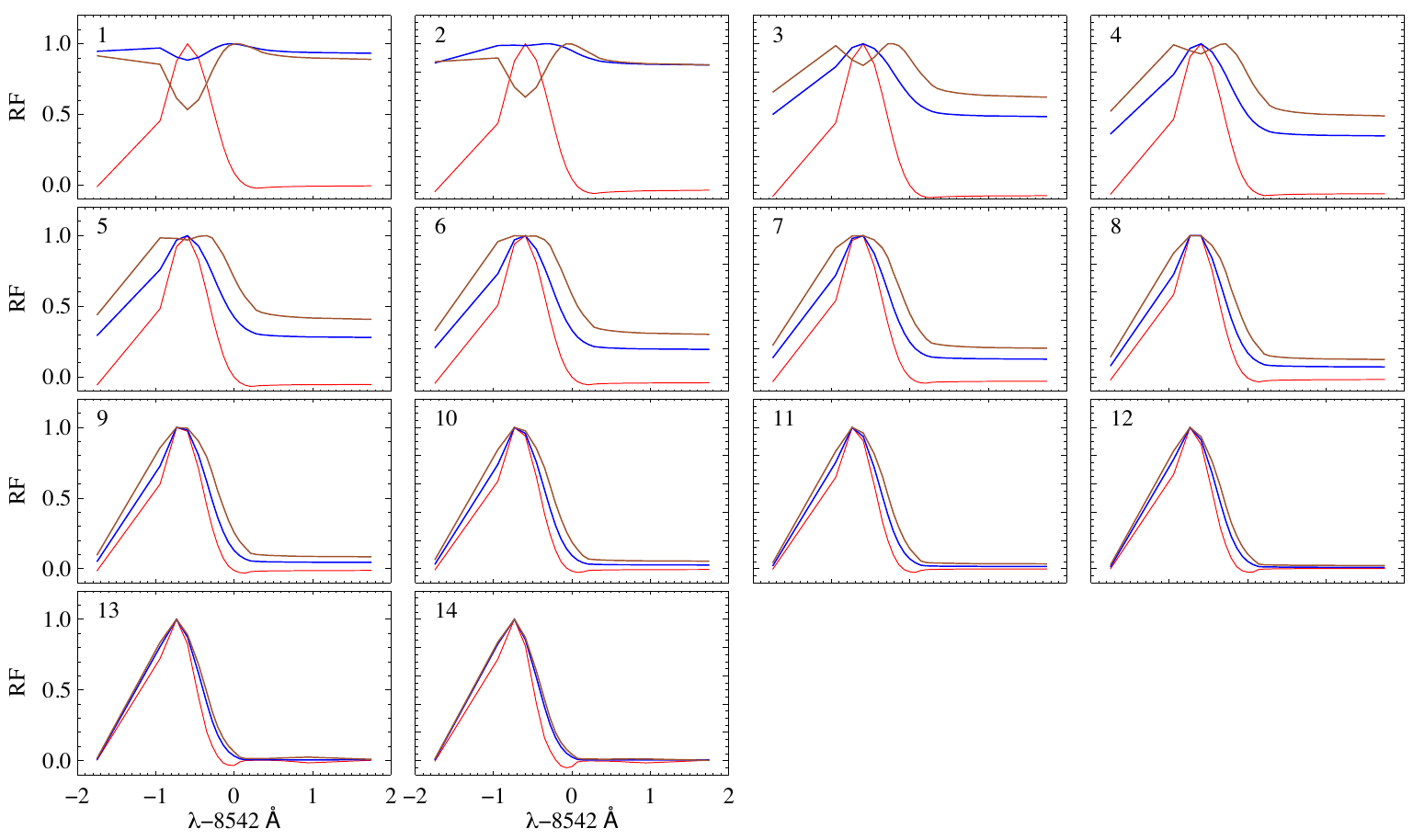}
\caption{Numerical Stokes {\it I} response functions of the
  \cawav\ line for temperature (red), spicule thickness (brown), and
  mass density (blue) for the blue component of spicule model
  normalized to their maximum values.} 
\label{fig7}
\end{figure}

As shown in Figure~\ref{fig6} the mass and electron densities
of the spicule plasma decrease exponentially as a function of
height. 
We note that the analysed structure 
bears many similarities 
and dissimilarities with conventional spicules making 
its unequivocal classification substantially difficult within current taxonomy 
of diverse small-scale jets. The structure has a similar width, length and line 
profiles to the spicules analysed in \citet{Beckers1968,Beckers1972}, 
providing the main motivation to interpret it as spicule. 
However, density obtained through the inversion makes this structure more 
similar to the larger scale chromospheric jets (macrospicule/surges) rather than typical spicules.
The values of the electron density $n_{\rm e}$
at four different times near the base of the spicule at a height of
$z\sim2$\,Mm is around $10^{13}$\,\cmcub.
\citet{Beckers1972} derived $n_{\rm e}\sim1.6\times 10^{11}$\,\cmcub\ 
at height of $z\sim2$\,Mm, around 60 times less than our estimates.  
Furthermore, \citet{Beckers1972} estimates that $n_{\rm e}$
decreases by 20 times at the top of the spicule at height of
$\sim 10$\,Mm, whereas our models show that the mass and
 electron densities decrease by $\sim100$
times at height of $z\sim9$\,Mm for the first three models
(Figure~\ref{fig6}: top left panel). The model at a late phase
of the spicule (at 08:41\,UT, around 2 min before the
disappearance) shows a sharper drop in density and
pressure. \citet{Alissandrakis1973} performed multiline spectroscopy
of 36 spicules and found that the average values of the electron
density at a height of 5\,400\,km is around
$6\times10^{10}$\,\cmcub. Our models show that the $n_{\rm e}$ at the
same height varies between $1 - 3\times10^{12}$\,\cmcub\, for
the four different models (Figure~\ref{fig6}: top left
panel).

The density and pressure scale heights of spicules measured from our
diagnostics are in the range of $ \Lambda\sim1\,000 - 2\,100$\,km over
the $\sim 2-9$\,Mm length projected to the surface normal
(Figure~\ref{fig6}: top left panel).  This is higher than some previous 
estimates of the scale heights in limb spicules using eclipse flash spectra
\citep{Makita2003} and magneto-seismology \citep{Verth2011}.  Density
scale heights derived by \citet{Bjolseth2008} from \ion{Ca}{2}\,H intensities
in spicules vary between $\sim2-8$\,Mm (see Figure 4.10 of
\citet{Bjolseth2008}).  However, \citet{Bjolseth2008} assumes that the
source function is constant along the spicule, and hence the scale
height of the intensity is defined only by the density
stratification. In our case, the wavelength-integrated
\ca\ intensities (not shown) decrease by a factor of $\sim10 - 16$ in
the range $2 - 8$\,Mm along the spicule, much slower than density and
pressure (Figure~\ref{fig6}: top left panel). Estimated
density/pressure scale heights are also much higher than the scale height of $\sim 290$\,km
of gas with comparable temperature in hydrostatic equilibrium.  This
is not surprising as the observed spicule displays a complex dynamic
behavior 
during its lifetime. 
Inversions distinguish two LoS velocity components along the
spicule.

The double-peak \cawav\, profiles detected along the jet axis are likely due to the
unwinding of highly twisted field lines rooted in magnetic topology with opposite polarities.
Unwinding twists can appear as a spinning motion of the plasma strands as found in rotating chromospheric jets \citep[see e.g.,][]{Shibata1986,Canfield1996,Liu2009,Liu2011,Pariat2015}.
The line-of-sight superposition of such strands may result in the double-peak profiles observed along the jet axis. 
\cite{Liu2009} has shown the observations of reconnection-driven jet, which involves untwisting helical threads rotating about the axis of a single large cylinder 
and shedding magnetic helicity into the upper atmosphere. 
They proposed the model in which flux emergence in an open-field region leads to magnetic reconnection, forming a jet and fan-spine topology. 
The twists in the reconnected open field tend to unwind themselves and drive these field lines to rotate \citep{Liu2011}.
The spicule analysed in this paper resembles much the fan-spine magnetic reconnection topology resulting from flux emergence studied in \cite{Liu2011}
and displays a spinning motion during its lifetime.

We note that the bi-directional flows along the spicular jet with parallel field lines in unipolar topology can also produce centrally reversed profiles of spicules. 
An observation of bi-directional flows at tops and bases of spicules is reported in \cite{Pasachoff2009}. 
Signature of counterstreams along a simulated macrospicule can be found in \cite{Murawski2011} for vertical or oblique magnetic field. 
However, \cite{Murawski2011} consider a localized velocity pulse as a driver of macrospicule, whereas the Y-shaped base of the spicule analysed in this study 
suggests the magnetic reconnection for the formation mechanism  (Figure 1). 
Furthermore, there is no clear and obvious evidence of the bi-directional flows in the observed spicule. 
Therefore, double-peak profiles of the spicule presented in this study is less likely to be a manifestation of counterstreaming or bi-directional flow.

The fact
  that the two spicule components have different 
  amplitudes of the maximum intensity could be due to the
  different filling factors within a resolution element.
  Indeed, inversions with height-dependent filling factors for
  red and blue components provided the best fit with the
  observations.  
  The exact
  relation of the spicule spectra with the dynamic motion and its
  transverse structure will be studied in a future paper where we plan
  to perform full 2D analysis and inversions of our observations.

The optical thickness $\tau$ in the \ca\ line center, i.e.
  where opacity reaches maximum, is a few hundreds at the spicule
  bottom and decreases exponentially toward its top. Over the main
  part of the spicule the $\tau$ is well above one, thus classifying
  the spicule plasma as optically thick in the \ca\ line
  center. \citet{Kralletal1976} performed theoretical calculations of
  the spicule optical thickness $\tau_{8542}$ in the \ca\ line center
  for the temperatures and electron densities ranging from 10\,000\,K
  to 20\,000\,K and from $2\times10^{10}$\,\cmcub\ to
  $1\times10^{11}$\,\cmcub, respectively, but inferring $\tau_{8542}$
  mostly of the order of $1\times10^{-2}-1\times10^{-3}$.
  They assumed the spicule thickness of 1\,425\,km (see page 100 therein), 
  thus larger than the thickness of 1\,000\,km employed here (Table~\ref{tab}).
  The high optical thickness could be related to the measured high density of spicule along its length.

We note that there are only limited number of radiative transfer
  codes that can deal with the off-limb emissions of the structures
  such as spicules.  One of the most advanced is non-LTE
  inversion code HAZEL \citep{Asensio2008}, which may be
    of interest for other similar investigations. We have
    decided to employ NICOLE for this work because of some unique
    features that are very important in this context, such as the
    ability to calculate  \ca\ populations in an optically thick
    medium with radiative transfer. 

Generally, the main problem of inversion techniques is the possible
non-uniqueness of the obtained solutions due to the degeneracy between
the different parameters. We have investigated the degeneracy between
temperature, density, and thickness of the spicule in the models
obtained through inversion by analyzing the response functions of the
intensity to the perturbations of these parameters. The results presented
in Figure~\ref{fig7} indicate that profiles 8 to 14 have a strong
degeneracy among the three parameters. This means that we could obtain
exactly the same solution with different combinations of those three
parameters. Therefore, if we were fitting those profiles individually,
we would have no way to obtain a unique solution. However, the
degeneracy is broken as the response functions are not the same for
the profiles of the bottom and middle parts of the spicule (see
profiles 1 to 7 in Figure~\ref{fig7}). As long as we fit all of the
profiles simultaneously the models obtained are free of this
degeneracy and the results are reasonably reliable.

We note that the spectral diagnostics presented here are applicable to
the local plasma parameters of a particular spicule and they may not
necessarily be considered as typical of all spicular
structures. However, it has been demonstrated that inversions can be
successfully applied to spicule spectra at multiple height
positions. With a wealth of statistics of semi-empirical spicule
models, we might gain a better understanding of chromospheric
spicules, their properties, and dynamics.
 
To our knowledge, this is the first spectral diagnostics of the
physical parameters of a spicule through line profile fitting
with a non-LTE inversion code.  The results are very encouraging for
future semi-empirical modeling of spicules using inversion techniques
and high-resolution spectropolarimetric observations with present and
future ground-based observing facilities.

\acknowledgments

The research leading to these results has received funding from the
S\^{e}r Cymru II scheme, part-funded by the European Regional
Development Fund through the Welsh Government, and STFC grant
ST/S000518/1 to Aberystwyth University.  The work of D.K. was
supported by Georgian Shota Rustaveli National Science Foundation
project FR17\_\ 323.  H.S.N. acknowledges support from the
Spanish Ministry of Economy and Competitivity through project
AYA2014-60476-P and PGC2018-102108-B-I00.  The Swedish 1-m Solar Telescope is operated on the
island of La Palma by the Institute for Solar Physics of Stockholm
University in the Spanish Observatorio del Roque de los Muchachos of
the Instituto de Astrof\'{\i}sica de Canarias. The Institute for Solar
Physics is supported by a grant for research infrastructures of
national importance from the Swedish Research Council (registration
number 2017-00625). J.K. acknowledges the project VEGA 2/0048/20.
R.O. acknowledges support from the Spanish Ministry of
Economy and Competitiveness (MINECO) and FEDER funds through project
AYA2017-85465-P. Ellie Nicholson is gratefully acknowledged for language corrections.

\facility{SST(CRISP), NICOLE}

\bibliography{dkuridze_etal}{}

\begin{thebibliography}{}
\expandafter\ifx\csname natexlab\endcsname\relax\def\natexlab#1{#1}\fi
\providecommand{\url}[1]{\href{#1}{#1}}
\providecommand{\dodoi}[1]{doi:~\href{http://doi.org/#1}{\nolinkurl{#1}}}
\providecommand{\doeprint}[1]{\href{http://ascl.net/#1}{\nolinkurl{http://ascl.net/#1}}}
\providecommand{\doarXiv}[1]{\href{https://arxiv.org/abs/#1}{\nolinkurl{https://arxiv.org/abs/#1}}}

\bibitem[{{Alissandrakis}(1973)}]{Alissandrakis1973}
{Alissandrakis}, C.~E. 1973, \solphys, 32, 345, \dodoi{10.1007/BF00154947}

\bibitem[{{Alissandrakis} {et~al.}(1990){Alissandrakis}, {Tsiropoula}, \&
  {Mein}}]{Alissandrakis1990}
{Alissandrakis}, C.~E., {Tsiropoula}, G., \& {Mein}, P. 1990, \aap, 230, 200

\bibitem[{{Alissandrakis} {et~al.}(2018){Alissandrakis}, {Vial}, {Koukras},
  {Buchlin}, \& {Chane-Yook}}]{Alissandrikisetal2018}
{Alissandrakis}, C.~E., {Vial}, J.~C., {Koukras}, A., {Buchlin}, E., \&
  {Chane-Yook}, M. 2018, \solphys, 293, 20, \dodoi{10.1007/s11207-018-1242-4}

\bibitem[{{Asensio Ramos} {et~al.}(2008){Asensio Ramos}, {Trujillo Bueno}, \&
  {Land i Degl'Innocenti}}]{Asensio2008}
{Asensio Ramos}, A., {Trujillo Bueno}, J., \& {Land i Degl'Innocenti}, E. 2008,
  \apj, 683, 542, \dodoi{10.1086/589433}

\bibitem[{{Beck} {et~al.}(2016){Beck}, {Rezaei}, {Puschmann}, \&
  {Fabbian}}]{Becketal2016}
{Beck}, C., {Rezaei}, R., {Puschmann}, K.~G., \& {Fabbian}, D. 2016, \solphys,
  291, 2281

\bibitem[{{Beckers}(1964)}]{Beckers1964}
{Beckers}, J.~M. 1964, \apj, 140, 1339, \dodoi{10.1086/148038}

\bibitem[{{Beckers}(1968)}]{Beckers1968}
---. 1968, \solphys, 3, 367, \dodoi{10.1007/BF00171614}

\bibitem[{{Beckers}(1972)}]{Beckers1972}
---. 1972, \araa, 10, 73, \dodoi{10.1146/annurev.aa.10.090172.000445}

\bibitem[{{Bj\o lseth}(2008)}]{Bjolseth2008}
{Bj\o lseth}, S. 2008, Masters thesis, Univ. Oslo

\bibitem[{{Canfield} {et~al.}(1996){Canfield}, {Reardon}, {Leka}, {Shibata},
  {Yokoyama}, \& {Shimojo}}]{Canfield1996}
{Canfield}, R.~C., {Reardon}, K.~P., {Leka}, K.~D., {et~al.} 1996, \apj, 464,
  1016, \dodoi{10.1086/177389}

\bibitem[{{Cauzzi} {et~al.}(2008){Cauzzi}, {Reardon}, {Uitenbroek},
  {Cavallini}, {Falchi}, {Falciani}, {Janssen}, {Rimmele}, {Vecchio}, \&
  {W{\"o}ger}}]{Cauzzi2008}
{Cauzzi}, G., {Reardon}, K.~P., {Uitenbroek}, H., {et~al.} 2008, \aap, 480,
  515, \dodoi{10.1051/0004-6361:20078642}

\bibitem[{{Centeno} {et~al.}(2010){Centeno}, {Trujillo Bueno}, \& {Asensio
  Ramos}}]{Centenoetal2010}
{Centeno}, R., {Trujillo Bueno}, J., \& {Asensio Ramos}, A. 2010, \apj, 708,
  1579, \dodoi{10.1088/0004-637X/708/2/1579}

\bibitem[{{de la Cruz Rodr{\'\i}guez} {et~al.}(2015){de la Cruz
  Rodr{\'\i}guez}, {Hansteen}, {Bellot-Rubio}, \& {Ortiz}}]{delaCruz2015}
{de la Cruz Rodr{\'\i}guez}, J., {Hansteen}, V., {Bellot-Rubio}, L., \&
  {Ortiz}, A. 2015, \apj, 810, 145, \dodoi{10.1088/0004-637X/810/2/145}

\bibitem[{{de la Cruz Rodr{\'{\i}}guez} {et~al.}(2015){de la Cruz
  Rodr{\'{\i}}guez}, {L{\"o}fdahl}, {S{\"u}tterlin}, {Hillberg}, \& {Rouppe van
  der Voort}}]{delaCruzRodriguezetal2015}
{de la Cruz Rodr{\'{\i}}guez}, J., {L{\"o}fdahl}, M.~G., {S{\"u}tterlin}, P.,
  {Hillberg}, T., \& {Rouppe van der Voort}, L. 2015, \aap, 573, A40,
  \dodoi{10.1051/0004-6361/201424319}

\bibitem[{{de la Cruz Rodr{\'\i}guez} {et~al.}(2013){de la Cruz
  Rodr{\'\i}guez}, {Rouppe van der Voort}, {Socas-Navarro}, \& {van
  Noort}}]{delaCruz2013}
{de la Cruz Rodr{\'\i}guez}, J., {Rouppe van der Voort}, L., {Socas-Navarro},
  H., \& {van Noort}, M. 2013, \aap, 556, A115,
  \dodoi{10.1051/0004-6361/201321629}

\bibitem[{{de la Cruz Rodr{\'\i}guez} \& {van Noort}(2017)}]{delaCruz2017}
{de la Cruz Rodr{\'\i}guez}, J., \& {van Noort}, M. 2017, \ssr, 210, 109,
  \dodoi{10.1007/s11214-016-0294-8}

\bibitem[{{De Pontieu} {et~al.}(2004){De Pontieu}, {Erd{\'e}lyi}, \&
  {James}}]{dePontieuetal2004}
{De Pontieu}, B., {Erd{\'e}lyi}, R., \& {James}, S.~P. 2004, \nat, 430, 536,
  \dodoi{10.1038/nature02749}

\bibitem[{{de Pontieu} {et~al.}(2007){de Pontieu}, {McIntosh}, {Hansteen},
  {Carlsson}, {Schrijver}, {Tarbell}, {Title}, {Shine}, {Suematsu}, {Tsuneta},
  {Katsukawa}, {Ichimoto}, {Shimizu}, \& {Nagata}}]{dePontieu2007}
{de Pontieu}, B., {McIntosh}, S., {Hansteen}, V.~H., {et~al.} 2007, \pasj, 59,
  S655, \dodoi{10.1093/pasj/59.sp3.S655}

\bibitem[{{De Pontieu} {et~al.}(2014){De Pontieu}, {Title}, {Lemen}, {Kushner},
  {Akin}, {Allard}, {Berger}, {Boerner}, {Cheung}, {Chou}, {Drake}, {Duncan},
  {Freeland}, {Heyman}, {Hoffman}, {Hurlburt}, {Lindgren}, {Mathur}, {Rehse},
  {Sabolish}, {Seguin}, {Schrijver}, {Tarbell}, {W{\"u}lser}, {Wolfson},
  {Yanari}, {Mudge}, {Nguyen-Phuc}, {Timmons}, {van Bezooijen}, {Weingrod},
  {Brookner}, {Butcher}, {Dougherty}, {Eder}, {Knagenhjelm}, {Larsen},
  {Mansir}, {Phan}, {Boyle}, {Cheimets}, {DeLuca}, {Golub}, {Gates}, {Hertz},
  {McKillop}, {Park}, {Perry}, {Podgorski}, {Reeves}, {Saar}, {Testa}, {Tian},
  {Weber}, {Dunn}, {Eccles}, {Jaeggli}, {Kankelborg}, {Mashburn}, {Pust},
  {Springer}, {Carvalho}, {Kleint}, {Marmie}, {Mazmanian}, {Pereira}, {Sawyer},
  {Strong}, {Worden}, {Carlsson}, {Hansteen}, {Leenaarts}, {Wiesmann},
  {Aloise}, {Chu}, {Bush}, {Scherrer}, {Brekke}, {Martinez-Sykora}, {Lites},
  {McIntosh}, {Uitenbroek}, {Okamoto}, {Gummin}, {Auker}, {Jerram}, {Pool}, \&
  {Waltham}}]{DePontieu2014}
{De Pontieu}, B., {Title}, A.~M., {Lemen}, J.~R., {et~al.} 2014, \solphys, 289,
  2733, \dodoi{10.1007/s11207-014-0485-y}

\bibitem[{{Gingerich} {et~al.}(1971){Gingerich}, {Noyes}, {Kalkofen}, \&
  {Cuny}}]{Gingerich1971}
{Gingerich}, O., {Noyes}, R.~W., {Kalkofen}, W., \& {Cuny}, Y. 1971, \solphys,
  18, 347, \dodoi{10.1007/BF00149057}

\bibitem[{{Hansteen} {et~al.}(2006){Hansteen}, {De Pontieu}, {Rouppe van der
  Voort}, {van Noort}, \& {Carlsson}}]{Hansteenetal2006}
{Hansteen}, V.~H., {De Pontieu}, B., {Rouppe van der Voort}, L., {van Noort},
  M., \& {Carlsson}, M. 2006, \apjl, 647, L73, \dodoi{10.1086/507452}

\bibitem[{{He} {et~al.}(2009){He}, {Marsch}, {Tu}, \& {Tian}}]{Heetal2009}
{He}, J., {Marsch}, E., {Tu}, C., \& {Tian}, H. 2009, \apjl, 705, L217,
  \dodoi{10.1088/0004-637X/705/2/L217}

\bibitem[{{Krall} {et~al.}(1976){Krall}, {Bessey}, \&
  {Beckers}}]{Kralletal1976}
{Krall}, K.~R., {Bessey}, R.~J., \& {Beckers}, J.~M. 1976, \solphys, 46, 93,
  \dodoi{10.1007/BF00157556}

\bibitem[{{Kriginsky} {et~al.}(2020){Kriginsky}, {Oliver}, {Freij}, {Kuridze},
  {Asensio Ramos}, \& {Antolin}}]{Kriginsky2020}
{Kriginsky}, M., {Oliver}, R., {Freij}, N., {et~al.} 2020, \aap, 642, A61,
  \dodoi{10.1051/0004-6361/202038546}

\bibitem[{{Kuridze} {et~al.}(2015{\natexlab{a}}){Kuridze}, {Henriques},
  {Mathioudakis}, {Erd{\'e}lyi}, {Zaqarashvili}, {Shelyag}, {Keys}, \&
  {Keenan}}]{Kuridze2015}
{Kuridze}, D., {Henriques}, V., {Mathioudakis}, M., {et~al.}
  2015{\natexlab{a}}, \apj, 802, 26, \dodoi{10.1088/0004-637X/802/1/26}

\bibitem[{{Kuridze} {et~al.}(2017){Kuridze}, {Henriques}, {Mathioudakis},
  {Koza}, {Zaqarashvili}, {Ryb{\'a}k}, {Hanslmeier}, \& {Keenan}}]{Kuridze2017}
---. 2017, \apj, 846, 9, \dodoi{10.3847/1538-4357/aa83b9}

\bibitem[{{Kuridze} {et~al.}(2018){Kuridze}, {Henriques}, {Mathioudakis},
  {Rouppe van der Voort}, {de la Cruz Rodr{\'\i}guez}, \&
  {Carlsson}}]{Kuridze2018}
{Kuridze}, D., {Henriques}, V.~M.~J., {Mathioudakis}, M., {et~al.} 2018, \apj,
  860, 10, \dodoi{10.3847/1538-4357/aac26d}

\bibitem[{{Kuridze} {et~al.}(2013){Kuridze}, {Verth}, {Mathioudakis},
  {Erd{\'e}lyi}, {Jess}, {Morton}, {Christian}, \& {Keenan}}]{Kuridze2013}
{Kuridze}, D., {Verth}, G., {Mathioudakis}, M., {et~al.} 2013, \apj, 779, 82,
  \dodoi{10.1088/0004-637X/779/1/82}

\bibitem[{{Kuridze} {et~al.}(2015{\natexlab{b}}){Kuridze}, {Mathioudakis},
  {Sim{\~o}es}, {Rouppe van der Voort}, {Carlsson}, {Jafarzadeh}, {Allred},
  {Kowalski}, {Kennedy}, {Fletcher}, {Graham}, \& {Keenan}}]{Kuridze2015b}
{Kuridze}, D., {Mathioudakis}, M., {Sim{\~o}es}, P.~J.~A., {et~al.}
  2015{\natexlab{b}}, \apj, 813, 125, \dodoi{10.1088/0004-637X/813/2/125}

\bibitem[{{Langangen} {et~al.}(2008){Langangen}, {De Pontieu}, {Carlsson},
  {Hansteen}, {Cauzzi}, \& {Reardon}}]{Langangen2008}
{Langangen}, {\O}., {De Pontieu}, B., {Carlsson}, M., {et~al.} 2008, \apjl,
  679, L167, \dodoi{10.1086/589442}

\bibitem[{{Linsky} {et~al.}(1970){Linsky}, {Teske}, \&
  {Wilkinson}}]{Linskyetal1970}
{Linsky}, J.~L., {Teske}, R.~G., \& {Wilkinson}, C.~W. 1970, \solphys, 11, 374,
  \dodoi{10.1007/BF00153072}

\bibitem[{{Lites}(1983)}]{Lites1983}
{Lites}, B.~W. 1983, \solphys, 85, 193, \dodoi{10.1007/BF00148648}

\bibitem[{{Liu} {et~al.}(2009){Liu}, {Berger}, {Title}, \& {Tarbell}}]{Liu2009}
{Liu}, W., {Berger}, T.~E., {Title}, A.~M., \& {Tarbell}, T.~D. 2009, \apjl,
  707, L37, \dodoi{10.1088/0004-637X/707/1/L37}

\bibitem[{{Liu} {et~al.}(2011){Liu}, {Berger}, {Title}, {Tarbell}, \&
  {Low}}]{Liu2011}
{Liu}, W., {Berger}, T.~E., {Title}, A.~M., {Tarbell}, T.~D., \& {Low}, B.~C.
  2011, \apj, 728, 103, \dodoi{10.1088/0004-637X/728/2/103}

\bibitem[{{L{\"o}fdahl}(2002)}]{Lofdahl2002}
{L{\"o}fdahl}, M.~G. 2002, in Society of Photo-Optical Instrumentation
  Engineers (SPIE) Conference Series, Vol. 4792, Image Reconstruction from
  Incomplete Data, ed. P.~J. {Bones}, M.~A. {Fiddy}, \& R.~P. {Millane},
  146--155, \dodoi{10.1117/12.451791}

\bibitem[{{L{\"o}fdahl} {et~al.}(2018){L{\"o}fdahl}, {Hillberg}, {de la Cruz
  Rodriguez}, {Vissers}, {Scharmer}, {Hagfors Haugan}, \&
  {Fredvik}}]{Lofdahletal2018}
{L{\"o}fdahl}, M.~G., {Hillberg}, T., {de la Cruz Rodriguez}, J., {et~al.}
  2018, ArXiv e-prints.
\newblock \doarXiv{1804.03030}

\bibitem[{{L{\'o}pez Ariste} \& {Casini}(2005)}]{LopezAristeAndCasini2005}
{L{\'o}pez Ariste}, A., \& {Casini}, R. 2005, \aap, 436, 325,
  \dodoi{10.1051/0004-6361:20042214}

\bibitem[{{Makita}(2003)}]{Makita2003}
{Makita}, M. 2003, Publications of the National Astronomical Observatory of
  Japan, 7, 1

\bibitem[{{Matsuno} \& {Hirayama}(1988)}]{MatsunoHirayama1988}
{Matsuno}, K., \& {Hirayama}, T. 1988, \solphys, 117, 21,
  \dodoi{10.1007/BF00148569}

\bibitem[{{Morita} {et~al.}(2010){Morita}, {Shibata}, {UeNo}, {Ichimoto},
  {Kitai}, \& {Otsuji}}]{Morita2010}
{Morita}, S., {Shibata}, K., {UeNo}, S., {et~al.} 2010, \pasj, 62, 901,
  \dodoi{10.1093/pasj/62.4.901}

\bibitem[{{Morton}(2014)}]{Morton2014}
{Morton}, R.~J. 2014, \aap, 566, A90, \dodoi{10.1051/0004-6361/201423718}

\bibitem[{{Murawski} {et~al.}(2011){Murawski}, {Srivastava}, \&
  {Zaqarashvili}}]{Murawski2011}
{Murawski}, K., {Srivastava}, A.~K., \& {Zaqarashvili}, T.~V. 2011, \aap, 535,
  A58, \dodoi{10.1051/0004-6361/201117589}

\bibitem[{{Nelson} {et~al.}(2019){Nelson}, {Freij}, {Bennett}, {Erd{\'e}lyi},
  \& {Mathioudakis}}]{Nelsonetal2019}
{Nelson}, C.~J., {Freij}, N., {Bennett}, S., {Erd{\'e}lyi}, R., \&
  {Mathioudakis}, M. 2019, \apj, 883, 115, \dodoi{10.3847/1538-4357/ab3a54}

\bibitem[{{Nishizuka} {et~al.}(2011){Nishizuka}, {Nakamura}, {Kawate}, {Singh},
  \& {Shibata}}]{Nishizuka2011}
{Nishizuka}, N., {Nakamura}, T., {Kawate}, T., {Singh}, K.~A.~P., \& {Shibata},
  K. 2011, \apj, 731, 43, \dodoi{10.1088/0004-637X/731/1/43}

\bibitem[{{Orozco Su{\'a}rez} {et~al.}(2015){Orozco Su{\'a}rez}, {Asensio
  Ramos}, \& {Trujillo Bueno}}]{OrozcoSuarezetal2015}
{Orozco Su{\'a}rez}, D., {Asensio Ramos}, A., \& {Trujillo Bueno}, J. 2015,
  \apjl, 803, L18, \dodoi{10.1088/2041-8205/803/2/L18}

\bibitem[{{Pariat} {et~al.}(2015){Pariat}, {Dalmasse}, {DeVore}, {Antiochos},
  \& {Karpen}}]{Pariat2015}
{Pariat}, E., {Dalmasse}, K., {DeVore}, C.~R., {Antiochos}, S.~K., \& {Karpen},
  J.~T. 2015, \aap, 573, A130, \dodoi{10.1051/0004-6361/201424209}

\bibitem[{{Pasachoff} {et~al.}(2009){Pasachoff}, {Jacobson}, \&
  {Sterling}}]{Pasachoff2009}
{Pasachoff}, J.~M., {Jacobson}, W.~A., \& {Sterling}, A.~C. 2009, \solphys,
  260, 59, \dodoi{10.1007/s11207-009-9430-x}

\bibitem[{{Pereira} {et~al.}(2014){Pereira}, {De Pontieu}, {Carlsson},
  {Hansteen}, {Tarbell}, {Lemen}, {Title}, {Boerner}, {Hurlburt}, {W{\"u}lser},
  {Mart{\'\i}nez-Sykora}, {Kleint}, {Golub}, {McKillop}, {Reeves}, {Saar},
  {Testa}, {Tian}, {Jaeggli}, \& {Kankelborg}}]{Pereira2014}
{Pereira}, T.~M.~D., {De Pontieu}, B., {Carlsson}, M., {et~al.} 2014, \apjl,
  792, L15, \dodoi{10.1088/2041-8205/792/1/L15}

\bibitem[{{Pietarila} {et~al.}(2007){Pietarila}, {Socas-Navarro}, \&
  {Bogdan}}]{Pietarila2007}
{Pietarila}, A., {Socas-Navarro}, H., \& {Bogdan}, T. 2007, \apj, 663, 1386,
  \dodoi{10.1086/518714}

\bibitem[{{Ramelli} {et~al.}(2006){Ramelli}, {Bianda}, {Merenda}, \& {Trujillo
  Bueno}}]{Ramellietal2006}
{Ramelli}, R., {Bianda}, M., {Merenda}, L., \& {Trujillo Bueno}, T. 2006, in
  Astronomical Society of the Pacific Conference Series, Vol. 358, Solar
  Polarization 4, ed. R.~{Casini} \& B.~W. {Lites}, 448

\bibitem[{{Ramelli} {et~al.}(2005){Ramelli}, {Bianda}, {Trujillo Bueno},
  {Merenda}, \& {Stenflo}}]{Ramellietal2005}
{Ramelli}, R., {Bianda}, M., {Trujillo Bueno}, J., {Merenda}, L., \& {Stenflo},
  J.~O. 2005, in ESA Special Publication, Vol. 596, Chromospheric and Coronal
  Magnetic Fields, ed. D.~E. {Innes}, A.~{Lagg}, \& S.~A. {Solanki}, 82.1

\bibitem[{{Rouppe van der Voort} {et~al.}(2009){Rouppe van der Voort},
  {Leenaarts}, {de Pontieu}, {Carlsson}, \& {Vissers}}]{Voort2009}
{Rouppe van der Voort}, L., {Leenaarts}, J., {de Pontieu}, B., {Carlsson}, M.,
  \& {Vissers}, G. 2009, \apj, 705, 272, \dodoi{10.1088/0004-637X/705/1/272}

\bibitem[{{Scharmer}(2006)}]{Scharmer2006}
{Scharmer}, G.~B. 2006, \aap, 447, 1111, \dodoi{10.1051/0004-6361:20052981}

\bibitem[{{Scharmer} {et~al.}(2008){Scharmer}, {Narayan}, {Hillberg}, {de la
  Cruz Rodriguez}, {L{\"o}fdahl}, {Kiselman}, {S{\"u}tterlin}, {van Noort}, \&
  {Lagg}}]{Scharmeretal2008}
{Scharmer}, G.~B., {Narayan}, G., {Hillberg}, T., {et~al.} 2008, \apjl, 689,
  L69, \dodoi{10.1086/595744}

\bibitem[{{Secchi}(1875)}]{Secchi1975}
{Secchi}, A. 1875, {Le Soleil} (Gauthier-Villars, Paris),
  \dodoi{10.3931/e-rara-14748}

\bibitem[{{Shibata} \& {Uchida}(1986)}]{Shibata1986}
{Shibata}, K., \& {Uchida}, Y. 1986, \solphys, 103, 299,
  \dodoi{10.1007/BF00147831}

\bibitem[{{Shibata} {et~al.}(2007){Shibata}, {Nakamura}, {Matsumoto}, {Otsuji},
  {Okamoto}, {Nishizuka}, {Kawate}, {Watanabe}, {Nagata}, {UeNo}, {Kitai},
  {Nozawa}, {Tsuneta}, {Suematsu}, {Ichimoto}, {Shimizu}, {Katsukawa},
  {Tarbell}, {Berger}, {Lites}, {Shine}, \& {Title}}]{Shibata2007}
{Shibata}, K., {Nakamura}, T., {Matsumoto}, T., {et~al.} 2007, Science, 318,
  1591, \dodoi{10.1126/science.1146708}

\bibitem[{{Shimojo} {et~al.}(2020){Shimojo}, {Kawate}, {Okamoto}, {Yokoyama},
  {Narukage}, {Sakao}, {Iwai}, {Fleishman}, \& {Shibata}}]{Shimojo2020}
{Shimojo}, M., {Kawate}, T., {Okamoto}, T.~J., {et~al.} 2020, \apjl, 888, L28,
  \dodoi{10.3847/2041-8213/ab62a5}

\bibitem[{{Skogsrud} {et~al.}(2015){Skogsrud}, {Rouppe van der Voort}, {De
  Pontieu}, \& {Pereira}}]{Skogsurd2015}
{Skogsrud}, H., {Rouppe van der Voort}, L., {De Pontieu}, B., \& {Pereira},
  T.~M.~D. 2015, \apj, 806, 170, \dodoi{10.1088/0004-637X/806/2/170}

\bibitem[{{Socas-Navarro} {et~al.}(2015){Socas-Navarro}, {de la Cruz
  Rodr{\'\i}guez}, {Asensio Ramos}, {Trujillo Bueno}, \& {Ruiz
  Cobo}}]{SocasNavarro2015}
{Socas-Navarro}, H., {de la Cruz Rodr{\'\i}guez}, J., {Asensio Ramos}, A.,
  {Trujillo Bueno}, J., \& {Ruiz Cobo}, B. 2015, \aap, 577, A7,
  \dodoi{10.1051/0004-6361/201424860}

\bibitem[{{Socas-Navarro} \& {Elmore}(2005)}]{SocasNavarroAndElmore2005}
{Socas-Navarro}, H., \& {Elmore}, D. 2005, \apjl, 619, L195,
  \dodoi{10.1086/428399}

\bibitem[{{Socas-Navarro} \& {Trujillo Bueno}(1997)}]{SocasNavarro1997}
{Socas-Navarro}, H., \& {Trujillo Bueno}, J. 1997, \apj, 490, 383,
  \dodoi{10.1086/304873}

\bibitem[{{Socas-Navarro} {et~al.}(2000){Socas-Navarro}, {Trujillo Bueno}, \&
  {Ruiz Cobo}}]{SocasNavarro2000}
{Socas-Navarro}, H., {Trujillo Bueno}, J., \& {Ruiz Cobo}, B. 2000, \apj, 530,
  977, \dodoi{10.1086/308414}

\bibitem[{{Sterling}(2000)}]{Sterling2000}
{Sterling}, A.~C. 2000, \solphys, 196, 79, \dodoi{10.1023/A:1005213923962}

\bibitem[{{Suematsu} {et~al.}(1995){Suematsu}, {Wang}, \&
  {Zirin}}]{Suematsu1995}
{Suematsu}, Y., {Wang}, H., \& {Zirin}, H. 1995, \apj, 450, 411,
  \dodoi{10.1086/176151}

\bibitem[{{Tei} {et~al.}(2020){Tei}, {Gun{\'a}r}, {Heinzel}, {Okamoto},
  {{\v{S}}t{\v{e}}p{\'a}n}, {Jej{\v{c}}i{\v{c}}}, \& {Shibata}}]{Tei2020}
{Tei}, A., {Gun{\'a}r}, S., {Heinzel}, P., {et~al.} 2020, \apj, 888, 42,
  \dodoi{10.3847/1538-4357/ab5db1}

\bibitem[{{Trujillo Bueno}(2010)}]{TrujilloBueno2010}
{Trujillo Bueno}, J. 2010, Astrophysics and Space Science Proceedings, 19, 118,
  \dodoi{10.1007/978-3-642-02859-5_9}

\bibitem[{{Trujillo Bueno} {et~al.}(2005){Trujillo Bueno}, {Merenda},
  {Centeno}, {Collados}, \& {Landi Degl'Innocenti}}]{TrujilloBuenoetal2005}
{Trujillo Bueno}, J., {Merenda}, L., {Centeno}, R., {Collados}, M., \& {Landi
  Degl'Innocenti}, E. 2005, \apjl, 619, L191, \dodoi{10.1086/428124}

\bibitem[{{Tsiropoula} \& {Schmieder}(1997)}]{Tsiropoula1997}
{Tsiropoula}, G., \& {Schmieder}, B. 1997, \aap, 324, 1183

\bibitem[{{Tsiropoula} {et~al.}(2012){Tsiropoula}, {Tziotziou}, {Kontogiannis},
  {Madjarska}, {Doyle}, \& {Suematsu}}]{Tsiropoula2012}
{Tsiropoula}, G., {Tziotziou}, K., {Kontogiannis}, I., {et~al.} 2012, \ssr,
  169, 181, \dodoi{10.1007/s11214-012-9920-2}

\bibitem[{{Tziotziou} {et~al.}(2003){Tziotziou}, {Tsiropoula}, \&
  {Mein}}]{Tziotziou2003}
{Tziotziou}, K., {Tsiropoula}, G., \& {Mein}, P. 2003, \aap, 402, 361,
  \dodoi{10.1051/0004-6361:20030220}

\bibitem[{{Uitenbroek}(1989)}]{Uitenbroek1989}
{Uitenbroek}, H. 1989, \aap, 213, 360

\bibitem[{{van Noort} {et~al.}(2005){van Noort}, {Rouppe van der Voort}, \&
  {L{\"o}fdahl}}]{vanNoortetal2005}
{van Noort}, M., {Rouppe van der Voort}, L., \& {L{\"o}fdahl}, M.~G. 2005,
  \solphys, 228, 191, \dodoi{10.1007/s11207-005-5782-z}

\bibitem[{{Verth} {et~al.}(2011){Verth}, {Goossens}, \& {He}}]{Verth2011}
{Verth}, G., {Goossens}, M., \& {He}, J.~S. 2011, \apjl, 733, L15,
  \dodoi{10.1088/2041-8205/733/1/L15}

\bibitem[{{Wootten} \& {Thompson}(2009)}]{WoottenThompson2009}
{Wootten}, A., \& {Thompson}, A.~R. 2009, IEEE Proceedings, 97, 1463,
  \dodoi{10.1109/JPROC.2009.2020572}

\bibitem[{{Yokoyama} \& {Shibata}(1995)}]{YokoyamaShibata1995}
{Yokoyama}, T., \& {Shibata}, K. 1995, \nat, 375, 42, \dodoi{10.1038/375042a0}

\bibitem[{{Zaqarashvili} \& {Erd{\'e}lyi}(2009)}]{Zaqarashvili2009}
{Zaqarashvili}, T.~V., \& {Erd{\'e}lyi}, R. 2009, \ssr, 149, 355,
  \dodoi{10.1007/s11214-009-9549-y}

\bibitem[{{Zaqarashvili} {et~al.}(2007){Zaqarashvili}, {Khutsishvili},
  {Kukhianidze}, \& {Ramishvili}}]{Zaqarashvili2007}
{Zaqarashvili}, T.~V., {Khutsishvili}, E., {Kukhianidze}, V., \& {Ramishvili},
  G. 2007, \aap, 474, 627, \dodoi{10.1051/0004-6361:20077661}

\end{thebibliography}
\bibliographystyle{aasjournal}

\end{document}